\documentclass[aps,prb,superscriptaddress,twocolumn,floatfix,10pt]{revtex4-2}%
    \usepackage{amsmath} 
    \usepackage{amssymb} 
    \usepackage[mathscr]{euscript} 
    \usepackage{color} 
    \usepackage{hyperref} \hypersetup{colorlinks=true,citecolor=blue,linkcolor=blue,urlcolor=blue} 
    \usepackage{graphicx} 
    \usepackage{siunitx} 
    \usepackage{orcidlink} 
    \usepackage{chemmacros}
 \usepackage{multirow}
    \usepackage{upgreek}
    \usepackage[normalem]{ulem}
    \usepackage{siunitx}
    \usepackage{float}

    \makeatletter 
    \def\@hangfrom@section#1#2#3{\@hangfrom{#1#2}#3}
    \def\@hangfroms@section#1#2{#1#2}
    \renewcommand\frontmatter@abstractwidth{\dimexpr\textwidth\relax} \makeatother 
    \makeatletter
    
    \usepackage{lipsum}
    
    \usepackage{epstopdf}
    \AppendGraphicsExtensions{.tif}
    
    \definecolor{pranab_green}{rgb}{0,0.39,0}
    \definecolor{pranab_red}{rgb}{0.85,0.23,0.11}
    \definecolor{pranab_blue}{rgb}{0.13,0.18,0.40}

    \newcommand\sgout{\bgroup\markoverwith{\textcolor{red}{\rule[0.5ex]{2pt}{0.4pt}}}\ULon} 

    \renewcommand{\epsilon}{\varepsilon}
    \renewcommand{\vec}[1]{\mathbf{#1}}
    \newcommand{\epar}{\epsilon_\parallel}
    \newcommand{\eperp}{\epsilon_\perp}

\begin{document}
    
    
    \title{Terahertz volume plasmon-polariton modulation
    in all-dielectric hyperbolic metamaterials}
    
    
    \author{Stefano~Campanaro\,\orcidlink{0009-0008-4207-6958}}\email{stefano.campanaro@unimore.it}
    \affiliation{Dipartimento di Scienze Fisiche, Informatiche e Matematiche, Universit\'a degli Studi di Modena e Reggio Emilia, I-41125 Modena, Italy}
    
    \author{Luca~Bursi\,\orcidlink{0000-0002-4530-0424}}\email{luca.bursi@unimore.it}
    \affiliation{Dipartimento di Scienze Fisiche, Informatiche e Matematiche, Universit\'a degli Studi di Modena e Reggio Emilia, I-41125 Modena, Italy}
    \affiliation{Istituto Nanoscienze CNR-NANO-S3, I-41125 Modena,
    Italy }
    
    \author{Stefano~Curtarolo\,\orcidlink{0000-0003-0570-8238}}\email{stefano@duke.edu}
    \affiliation{Department of Mechanical Engineering and Materials Science, Duke University, Durham, NC 27708, USA}
    \affiliation{Center for Extreme Materials, Duke University, Durham, NC 27708, USA}
    
    \author{Arrigo~Calzolari\,\orcidlink{0000-0002-0244-7717}}\email{arrigo.calzolari@nano.cnr.it}
    \affiliation{Istituto Nanoscienze CNR-NANO-S3, I-41125 Modena,
    Italy }
    \affiliation{Department of Mechanical Engineering and Materials Science, Duke University, Durham, NC 27708, USA}
    \affiliation{Center for Extreme Materials, Duke University, Durham, NC 27708, USA}

    
    \begin{abstract}
    \noindent
    The development of plasmonics and related applications in the terahertz range faces limitations due to the intrinsic high electron density of standard metals. All-dielectric systems are profitable alternatives, which allows for customized modulation of the optical response upon doping. Here we focus on plasmon-based hyperbolic metamaterials realized stacking doped  III-V semiconductors that have been shown to be optically active in the terahertz spectral region. By using a multi-physics  multi-scale theoretical approach, we unravel the role of doping and geometrical characteristics  (e.g., thickness, composition, grating) in the modulation of high-$k$ plasmon-polariton modes across the metamaterial.
    \end{abstract}
    
    \maketitle
    %
\section{Introduction}
Hyperbolic metamaterials (HMMs) are artificial systems characterized by extreme
optical anisotropy as they behave as metals in one direction and as dielectrics in the perpendicular direction \cite{Smolyaninov2018,Poddubny:2013cy}. This implies that their dispersion relation $[\omega(\vec{k})=costant]$
is geometrically described by an open-surface hyperboloid and not by a closed ellipsoid, as in the case of ordinary materials. 
As such, HMMs can sustain extraordinary electromagnetic (EM) traveling modes with high-$k$ wavevectors \cite{excitation2014},
whose propagation is fully described by classical electrodynamics, once the dielectric functions of the composing materials are known.
Usually, HMMs are realized through stacked alternation of metallic (e.g., Au, Ag, Cu, TiN) and dielectric (e.g., TiO$_2$, ZnO, Al$_2$O$_3$) layers, and
the high-$k$  modes are the Bloch modes of
the metal-dielectric superlattice. They are volume plasmon-polaritons (VPPs) able to propagate along sharp cones across the stack \cite{morandi2023} that arise from the resonant coupling of surface plasmon-polaritons (SPPs) at each metal/dielectric interface \cite{vpp_hmm}. 
These unique optical properties  make HMMs suitable candidates for many advanced optical
applications like super-resolution imaging \cite{ferrari}, plasmonic sensors \cite{Strangi16,Kumar_2020}, new stealth technologies \cite{Yin15,Narimanov10},  and thermal emitters \cite{Biehs2012,Guo13}.

The majority of HMMs realized in the last years
are hyperbolically active  in the near-infrared (IR) to ultraviolet (UV) spectral range (see e.g., Refs. \cite{sh,Naik2012_AZO_ZnO,Naik2014_TiN_superlattice,Nan2024_Ag_TiO2,Jen2023_TiO2_Ag,pnas.161308111,acsnano.8b06026,Lee21}). However, it would be of great technological interest having HMMs working in the
terahertz (THz) region, from 300 GHz to 10 THz. This would bridge the gap between electronics and 
optics, traditionally challenging for THz radiation generation, confinement,  and 
detection \cite{nr_semiconductor,boltasseva}. Terahertz technology holds substantial potential for
a broad range of emerging applications in the fields of non-destructive material inspection and quality control \cite{Karali2018}, medical diagnostics \cite{Gong27052020}, imaging \cite{Jansen:10}, defense and security \cite{DAVIES200818}.

Because of the intrinsic high electron density (e.g., $n_e\sim10^{22}-10^{23}$ cm$^{-3}$), 
plasmonic metals -- including noble metals \cite{C2CS35367A}, transition metal nitrides \cite{li2014,Catellani2017,CalzolariAdv2021}, and high entropy carbides \cite{calzolari2022,Divilov2025,curtarolo:art80} -- are not suited to achieve the THz region, having the (screen) plasmon resonance in the near-IR or visible range. The large carrier densities of metals also lead to significant ohmic losses and offer limited spectral tunability, further hindering their use in low-frequency HMM platforms. 
On the contrary, all-dielectric metamaterials represent a profitable alternative, where the conductive components are realized by doped semiconductors that have lower (e.g., $n_e\sim10^{18}-10^{20}$ cm$^{-3}$) and controllable electron densities. Doped semiconductors offer reduced ohmic losses, and allow for full compatibility with high-quality epitaxial growth. In particular, narrow-gap III-V semiconductors (such as InAs and InSb) are well suited for THz and mid-IR plasmonics \cite{seren2016,Tong2021,Aupiais2023,Duan23,Ning11}, thanks to their low effective masses, high mobility, and the strong tunability of their plasmonic response through doping \cite{Chochol2017}, temperature \cite{Kono2010}, and electric fields \cite{Sai2023}. Recent experimental reports \cite{sohr2018,sohr2021,Desouky2017} have demonstrated the possibility to realize hyperbolic metamaterials in the mid-IR and THz frequency range, by using III-V semiconductor-based multilayers, highlighting the technological potential of this materials platform.

More generally, while realization of all-dielectric metamaterials has been experimentally proved \cite{sohr2018,sohr2021,Desouky2017,Jahani16,Wei16,nn203406f,Liang19,Wong13}, their theoretical characterization remains more subtle. Indeed,
 the optical properties of semiconductors are system-specific and strongly depend on the  growth conditions (e.g., composition, doping, defect concentration, etc.). Thus, the dielectric function is obtained through experimental means, otherwise modeling  the high-$k$ modes in all-dielectric HMMs would be impossible. Furthermore, even when  experimental permittivities are available,  they cannot be directly transferred to predict the hyperbolic response of other, albeit similar, compositions. On the theoretical side, while a few reports evaluate the plasmonic properties of doped semiconductors trough simulations from first-principles  that take explicitly into account the role of  composition and doping modulation \cite{PhysRevB.94.115208,PhysRevLett.133.116402,jp5046035,Jung13}, the {\em ab initio} description of HMMs  remains limited to the homogenized medium treatment of the dielectric function \cite{GJERDE2023112199,PhysRevB.103.035425,s41467-017-00412-y,PhysRevB.103.035425}.  This effective model  does not include the effects of structural parameters (such as number of layers, thickness,  grating coupler) that are crucial for the characterization and the prediction of  the extraordinary waves  within the metamaterials.
 This highlights the need for new theoretical approaches that combine materials science, quantum optics, and electromagnetism, where: {\bf i.} the description of materials goes beyond the use of single empirical parameters (e.g., dielectric constant) tabulated for the reference bulk materials; and {\bf ii.} the structural setup and the interface with the external environment are explicitly considered.

In this article, we attempt to bridge the gap between the experimental observations and the microscopic understanding of HMMs, by exploring their intrinsic properties. We systematically investigate metamaterials composed entirely of binary III-V semiconductors, demonstrating: {\bf i.} their feasibility in realizing HMMs active in the THz, {\bf ii.} their optical modulation with doping and multilayer geometry. We also provide a comprehensive theoretical analysis of  VPP modes, offering insights into how these metamaterials can be engineered to achieve desired optical characteristics.  In particular, we implemented  a multiscale approach that combines the microscopic descriptions of materials with the macroscopic representation of traveling EM waves, integrating  atomistic first principles simulations, effective medium theory, photonic band structure analysis, and continuum EM techniques based on  scattering matrix method. 

This research aims to advance the understanding of HMMs design principles, by enriching the EM description of VPP modes with system-tailored materials properties. The resulting critical insights could drive the optimization of semiconductor-based HMMs for applications in areas such as optoelectronics, sensing, and thermal management in the THz region.

\section{Theoretical methods and computational details}
\label{sec:method}

\noindent{\bf Materials characterization.}
Atomistic simulations based on density functional theory (DFT)  were carried out to calculate the electronic structure and the complex dielectric functions of both undoped and doped III-V semiconductors. All simulations were performed using the DFT engine ({\em pw.x}) included within the Quantum ESPRESSO (QE) suite \cite{qe,qe2017}. PBE generalized gradient approximation \cite{PBE1996} was applied to the exchange-correlation functional.  Electron-ion interactions were described using norm-conserving pseudopotentials \cite{Hamann2013} from the  ONCVSP  library \cite{ONCVPSP}. Single-particle wavefunctions were expanded in planewaves with a kinetic energy cutoff of 120 Ry. 
A uniform mesh of  ($16\times16\times16$) $\mathbf{k}$-point was used to sample the Brillouin zone  in self-consistent DFT cycles, and a denser ($60\times60\times60$) $\mathbf{k}$-grid was exploited in non-self-consistent field calculations needed for optical properties. Lattice parameters were optimized using the variable-cell minimization method. All structures were fully relaxed by using total-energy-and-force optimization scheme, until forces per atom are less than 0.03 eV/ \AA. Undoped semiconductors were simulated in the primitive {\em fcc} cell, while Si-doped InAs systems were modeled with periodic supercells of different sizes (see Section Results). 
 The well-known DFT underestimation  of energy band gap was corrected by using a pseudo-hybrid Hubbard implementation of
DFT+$U$ within the Dudarev formulation \cite{PhysRevB.57.1505}, namely ACBN0 \cite{hubbard_dft,acbn0}, which involves the use of the effective parameter $U_{eff}=U-J$, where $U$ is the corrective Hubbard term and 
$J$ accounts
for the energy cost associated with the exchange interaction.  The ACBN0 approach is  implemented in the PAOFLOW code \cite{paoflow2018}, which operates in a joint
loop with the DFT-based executables of QE.
The calculated $U_{eff}$ values for each chemical species are collected in Table S1 of the Supporting Information (SI). 

The complex dielectric function $\hat{\epsilon}(\omega)=\epsilon_1+i\epsilon_2$ is evaluated using the code {\em epsilon.x}, also included in the QE suite. This code implements an independent particle formulation of the frequency-dependent ($\omega$) Drude-Lorentz model for solids \cite{Calzolari:2014gj,Colle2007}, which explicitly includes both intraband and interband  transitions between Bloch states, along with  Drude-like and Lorentz-like
relaxation terms, which account for the  finite lifetime of the electronic excitations and  the effects
of the dissipative electron scattering \cite{wooten}.

\begin{figure}[t!]
\centering
\includegraphics[width=0.5\textwidth]{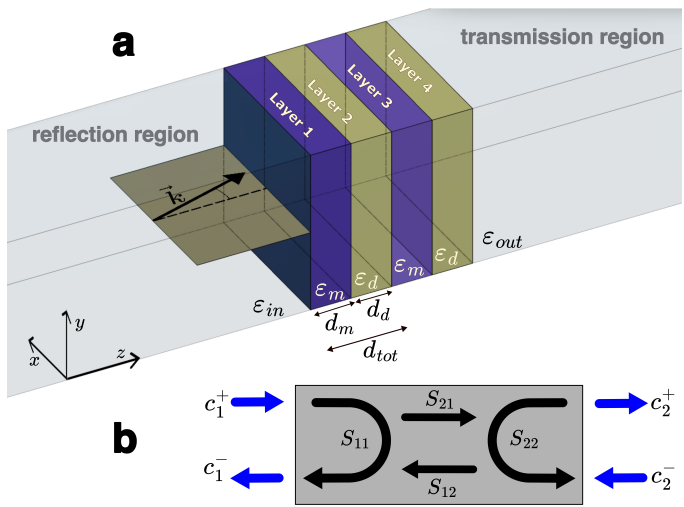}
\caption{(a) Schematic representation of a finite stacked metamaterial composed of alternating metallic ($m$) and dielectric ($d$) layers,  each characterized by its dielectric function $\varepsilon_j$ and thickness $d_j$, where $j=\{m,d\}$.  Gray regions indicate the embedding media for reflection and transmission, also characterized by the respective dielectric functions $\epsilon_{in}$ and $\epsilon_{out}$. The optical  axis is aligned along the {\em z} direction.
(b) Scattering matrix formulation for a single layer.
Forward $(c_1^+, c_2^+)$ and backward $(c_1^-, c_2^-)$ 
propagating wave amplitudes are related through the elements $S_{ij}$ of the scattering matrix.}
\label{fig:fig1}
\end{figure}

\noindent{\bf Effective medium theory (EMT).}
Effective medium theory \cite{em_multilayer}  is used to describe the overall optical properties of a periodically repeated (i.e., infinite) multilayer at the macroscopic level.  Under this approximation --- valid when the thickness of each constituent layer is much smaller than the wavelength of the probing radiation --- the effective dielectric functions in the directions parallel ($\epar$) and perpendicular ($\eperp$) to the optical  axis (Figure \ref{fig:fig1}a) are:
\begin{equation}\label{eps_perp}
\eperp= f_m \epsilon_m+f_d \epsilon_d \;,
\end{equation}
\begin{equation}\label{eps_par}
\epar=\frac{\epsilon_m\epsilon_d}{f_d\epsilon_m+f_m\epsilon_d}\;,
\end{equation}
where $\epsilon_m$ ($\epsilon_d$) is the complex dielectric function of the metallic (dielectric) material resulting from DFT+$U$ calculations;  
$f_j$ ($j=\{m,d\}$) is the layer filling factor defined as 
$f_j=d_j/d_{tot} $ where $d_{tot}=d_m+d_d$ is the thickness of the periodic unit cell (Figure \ref{fig:fig1}a). 
The macroscopic conditions for  VPP excitation and the angle between the extraordinary wave  and the optical axis are analysed through  
 the angular dielectric function  $\epsilon_{\varphi}$ and the angle $\Theta$, which are geometrical functions of $\epar$ and $\eperp$; see Sec. S1 of Supporting Information (SI).

\noindent{\bf Scattering matrix method (SMM).} 
The VPP propagation through a finite 3D multilayer  \cite{berreman, rumpf11} is described through the scattering matrix ({\em S}-matrix) \cite{born-wolf1999}, as implemented in the {\em AFLOW-EMERALD} code \cite{emerald}. 
By simulating finite multilayers with a defined number $\mathcal{N}$ of periods,  {\em AFLOW-EMERALD} provides the reflectivity, transmissivity, and absorption of EM waves across the stack, accounting for the interference effects at each interface and resonant phenomena due to the internal structure of the superlattice, as well as the coupling with the external environment. 
 For non-magnetic materials,  the scattering matrix is calculated for each layer and depends on the frequency of incident monochromatic radiation,
its planar wavevector ($k_x$),  and the dielectric permittivity ($\hat{\epsilon}$) that we obtained from DFT+$U$ simulations.
The matrix {\bf {\em S}} relates the amplitude coefficients of the forward ($c^+$) and backward ($c^-$) propagating components of the electric and magnetic fields at the left (1) and right (2) interfaces of the {\em n}-th layer, as illustrated in Figure \ref{fig:fig1}b. The reflectivity (R), transmissivity (T), and absorption (A) functions are calculated from the resulting EM fields being reflected, transmitted, and absorbed  through the entire multilayer \cite{emerald}; see Section S1 of SI for further details.

\noindent{\bf Photonic band structure (PBS).}
The  Bloch modes of periodic metal-dielectric superstructures are studied through the evaluation of the photonic band structure,
whose formulation is derived from photonic crystal theory \cite{AGIO2005286}. The 3D translational invariance of the periodic multilayers implies that the dielectric function is also invariant under translation by Bravais vectors $\mathbf{R}$, i.e., $\epsilon(\mathbf{r}+\mathbf{R})=\epsilon(\mathbf{r})$. This leads to the validity of the Bloch-Floquet theorem and the formation of allowed (photonic bands) and forbidden (photonic gaps) EM states.   
Photonic band structures are calculated with the {\em AFLOW-EMERALD} code \cite{emerald} by expanding  the electric and magnetic fields on a planewaves basis set, as detailed in Section S1 of SI. The dielectric function is derived from the first principles simulations of the composing bulk materials.

\section{Results}
The theoretical approaches employed in this work operate at complementary levels and together provide a coherent multiscale description of the metamaterials. EMT, informed by the material dielectric functions computed from first principles, offers a macroscopic picture of an infinite multilayer, described as a single, homogenized medium, and predicts the spectral regions in which the system behaves as Type-I, Type-II HMM, or as a dielectric. SMM extends the analysis to finite stacks, explicitly accounting for interface reflections, grating-assisted momentum matching, and coupling to the external environment, thus enabling the description of the propagation of EM waves through realistic metamaterials. Finally, PBS analysis identifies the allowed Bloch modes of the periodically repeated multilayer and rationalizes which  modes can be excited under the boundary conditions imposed by the finite-thickness stack and the matching coupler. Here, we considered ideal interfaces, this approximation can be systematically relaxed at different levels of the multiscale framework. The hierarchical use of EMT, SMM, and PBS enables us to connect material properties with propagating VPP modes in tailor-made metamaterial structures.

\noindent{\bf Optical proprieties of multilayer components.}
Along the lines of the experimental structures described in Refs. \cite{sohr2018,sohr2021}, we have modeled the HMMs as binary superlattices composed of 
alternating layers of dielectric (III-V semiconductors) and metallic (Si-doped InAs) materials, stacked along the $z$ axis. A
schematic representation of an all-dielectric HMM multilayer is shown in
Figure \ref{fig:fig1}a. In the experiments of reference, the range of interest of
the electromagnetic radiation was $\lambda \sim (2-20) \mu$m (i.e., THz to mid-IR range)
and the thickness of the building layers within the stack was of the order of 100 nm. Since the wavelength of the incoming
radiation is much greater than the thickness of the individual components, the homogenization assumptions \cite{em_multilayer} are largely satisfied, quantum confinement 
effects can be neglected, and the complex dielectric functions of the single bulk materials are sufficient to fully characterize the optical response of the overall metamaterials.

As first step, we evaluated electronic and optical properties of
the nine III-V binary semiconductors (namely,
AlP, AlAs, AlSb, GaP, GaAs, GaSb, InP, InAs, and InSb) in the {\em fcc} zincblend  phase. 
The resulting lattice parameters ($a_0$) and the  band gap values ($E_g$) are summarized in Table
S1 of SI. The results  indicate that the DFT+$U$ approach correctly reproduces the semiconducting behavior ($E_g>0$) of all the systems,
including narrow-gap semiconductors, such as InAs, that result semimetallic ($E_g<0$) in standard DFT calculations \cite{PhysRevB.80.035203}.
The band gap underestimation 
is significantly reduced with the DFT+$U$ correction, albeit a residual underestimation remains for all systems.  
The corresponding band structure plots (see Sec. S2 of SI) well agree with theoretical results obtained by using higher-level DFT approaches, 
such as hybrid functionals, or many-body GW approximation \cite{PhysRevB.82.205212,Durand14}. 
The frequency-dependent dielectric functions of the III-V
semiconductors have been
calculated  by using
a single-particle Drude-Lorentz approach \cite{grosso_solid}. The results of the simulations very well concur with previous theoretical calculations \cite{PhysRevB.71.045202} and with experimental data \cite{epsm_AlP, epsm_AlAs, epsm_AlSb}. The results for the nine III-V compounds are collected in Section S3 of SI.
All the main spectral features of both $\epsilon_1$ and $\epsilon_2$ (i.e., peak energies and peak intensities)  are accurately reproduced for all systems. 
Yet, with respect
to the experimental data, we observe a small but systematic rigid redshift in the inter-band optical absorption threshold, which is reminiscent of the residual underestimation of the simulated energy band gap discussed above. 

\begin{figure*}[!htb]
    \centering
    \includegraphics[width=0.8\textwidth]{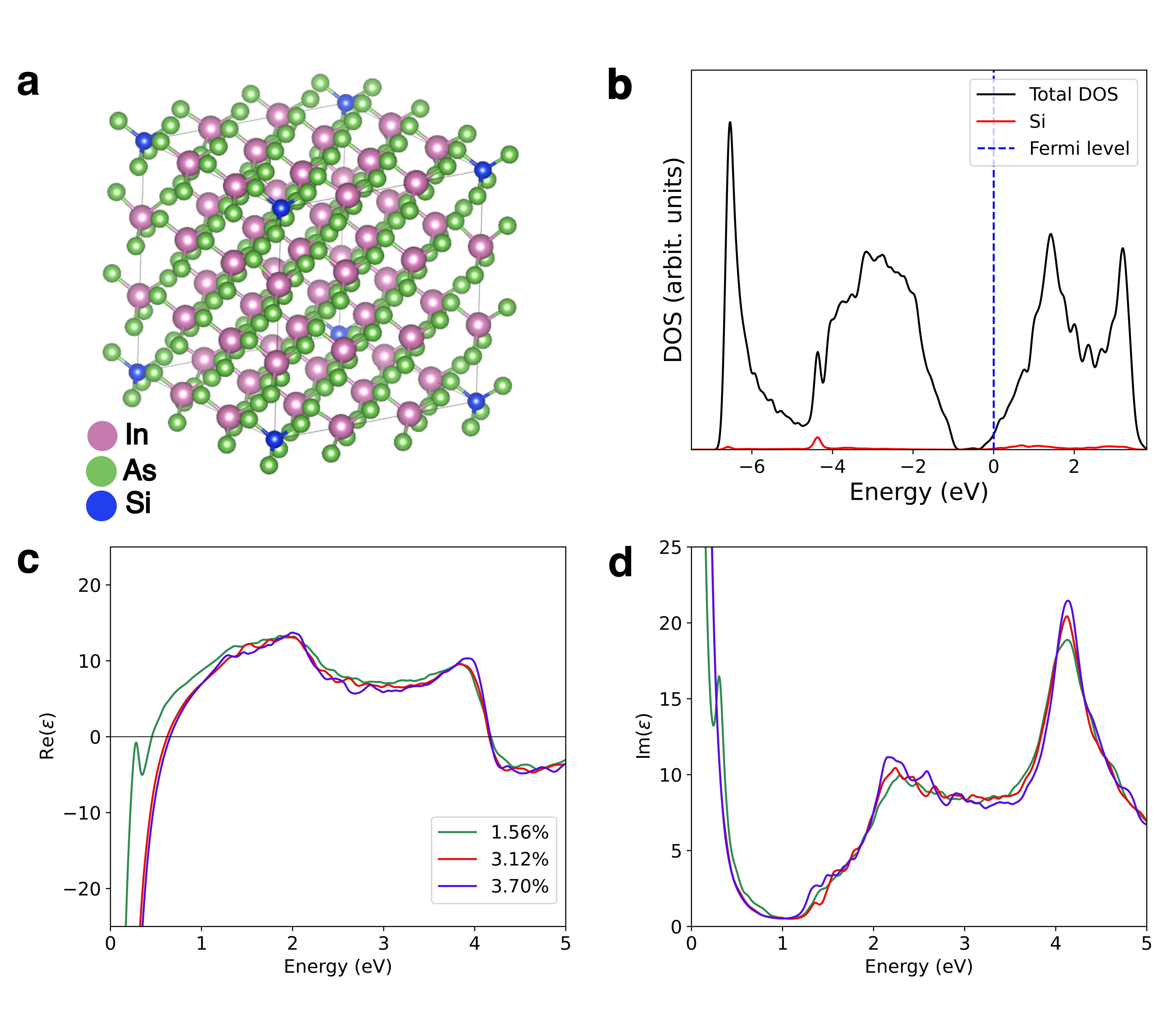}
	\caption{\small a) Atomic structure of Si:InAs supercell. b) Total (black line) and Si-projected (red line) density of states (DOS) of Si:InAs (model {\bf 2}). c) Real and d) imaginary parts of the dielectric function of Si:InAs as a function of Si concentration. }
	\label{fig:fig2}
\end{figure*}

In line with the experimental case \cite{sohr2018,sohr2021}, the metallic component in our multilayers is Si-doped InAs (Si:InAs) that we simulated at different Si concentrations in order to unravel
the role of free charge density on the plasmon-polariton properties across the metamaterials.  Si:InAs is modeled by  including In-substitutional Si atoms in extended
InAs supercells, within the virtual crystal approximation (Figure \ref{fig:fig2}a). The doping
concentration is defined as $c=N_{Si}/N_{In}$,
where $N_{Si}$ is the number of Si dopant atoms, and $N_{In}$ is the total number of In atoms in an equivalent undoped supercell.
In the experimental references \cite{sohr2018,sohr2021}, doping concentration is determined in terms of
the resulting free electron density $n_{el} \sim 10^{20}$ cm$^{-3}$, which corresponds to a nominal Si concentration $c\simeq2.2\%$.
Here, we generate three models (labeled {\bf 1}, {\bf 2}, and {\bf 3}) at varying doping concentrations in the range $1.6-3.7\%$, which includes the experimental case.
The details relative to concentration and cell size are summarized in Table \ref{tab:tab1}.
After full atomic relaxation, the results indicate that Si imparts a slight local distortion in the
structure that is almost insensitive to the doping concentration.

\begin{table}[!b]
\caption{\small Si-dopant concentration ($c$), supercell lattice parameter ($a_0$),  Si-In and Si-As  nearest neighbors distances for the three Si:InAs models described in the text. $\Delta E_F$ is the energy difference between the Fermi level ($E_F$) and the bottom of the conduction band; $E_p$ is the plasmon energy.}
\centering
\begin{tabular}{ccccccc}
\hline
Model   &$c$ (\%) &   $a_0$(\AA) & Si-In(\AA) & Si-As(\AA) & $\Delta E_F$(eV) & $E_p$(eV) \\ 
\hline
{\bf 1}&1.56\%  & 24.23 & 4.24 & 2.41&0.12&0.46\\
{\bf 2}&3.12\% & 18.17 & 4.24 & 2.41 &0.21&0.62\\
{\bf 3}&3.70\%  & 12.12 & 4.24 & 2.41 &0.25&0.65\\ 
\hline
\end{tabular}
\label{tab:tab1}
\end{table}
%
As expected, the ground state electronic structure of all models corresponds to an $n$-type degenerate conductor, as shown in Figure \ref{fig:fig2}b, for the case of medium-doping model {\bf 2}.  The density of states (DOS) of Si:InAs exhibits all the spectral
features of undoped InAs. Yet, the pristine energy gap of the host
 is slightly reduced in the doped system, while  no Si-derived  peaks
appear in the gap region. Si donates its extra {\em 3p} electron to the host conduction band, without changing much the shape and
the curvature of the conduction band minimum. As a consequence, doping causes a shift of the Fermi level into the conduction band and the
system acquires a metal-like behavior.
The other two models show very similar DOS spectra, which differ
only in the position of the Fermi level with respect to the bottom of the
conduction band ($\Delta E_F$ in Table \ref{tab:tab1}) that corresponds to a different injected charge,
which increases with Si concentration. By inverting the Drude formula in the parabolic approximation, it is possible to estimate, {\em a posteriori}, the effective free electron density generated by doping. The results are of the order of $n_e \sim 10^{19}$ cm$^{-3}$ for model {\bf 1}, and $n_e \sim 10^{20}$ cm$^{-3}$ for models {\bf 2} and {\bf 3}, slightly lower than the nominal values but still in agreement with the experimental data \cite{sohr2018,sohr2021}.

The dielectric functions of Si:InAs configurations have been calculated with the same computational procedure
and the same parameters detailed above. Figure \ref{fig:fig2} shows the resulting plots for the real (panel c) and imaginary (panel d) parts of the dielectric functions as a function of the doping concentration.
While the doping only marginally affects the optical properties above $E\sim$ 2 eV,
it completely changes the dielectric function in the low-energy part of the spectrum.
At low frequencies (i.e., $E\to 0$), the optical properties of Si:InAs  differ from the original InAs,
being  similar to a simple metal. 
$E_p$ 
identifies the energy at which the real part of the dielectric function changes sign $[\epsilon_1(E_p)=0]$, i.e., the energy at which the system undergoes an optical switch from metal- to dielectric-like. This is a key parameter in the
characterization of the hyperbolic properties of the metamaterials (see below).
In all  Si:InAs models,  $\epsilon_2(E_p)$ also is negligible (Figure \ref{fig:fig2}d). This corresponds to
the possibility of exciting a collective electronic oscillation of a reduced
 part of the free-electron density, known as a screened plasmon
resonance in the  THz range. In this case, $E_p$ represents the THz plasmon energy for the system.  
While  for pure metals, $E_p$ is an intrinsic property that cannot be changed, for doped-semiconductors the plasmon energy
can be tuned by controlling the electron density, i.e., by changing the doping conditions (Table \ref{tab:tab1}). 
Finally, it is worth noticing that the permittivity for low-doping concentration model ({\bf 1}) exhibits an extra peak at $\sim 0.3 $ eV, absent in the other two doping models. This structure arises from residual interband transitions in lightly doped InAs, occurring just above the fundamental gap. As the doping level increases, the Fermi level moves deeper into the conduction band, the Pauli blocking progressively suppresses these transitions, and the corresponding absorption feature naturally disappears at higher electron densities. The presence of this spectral feature does not change the plasmonic character of the Si:InAs model.

\begin{figure}[t]
\centering
\includegraphics[width=0.5\textwidth]{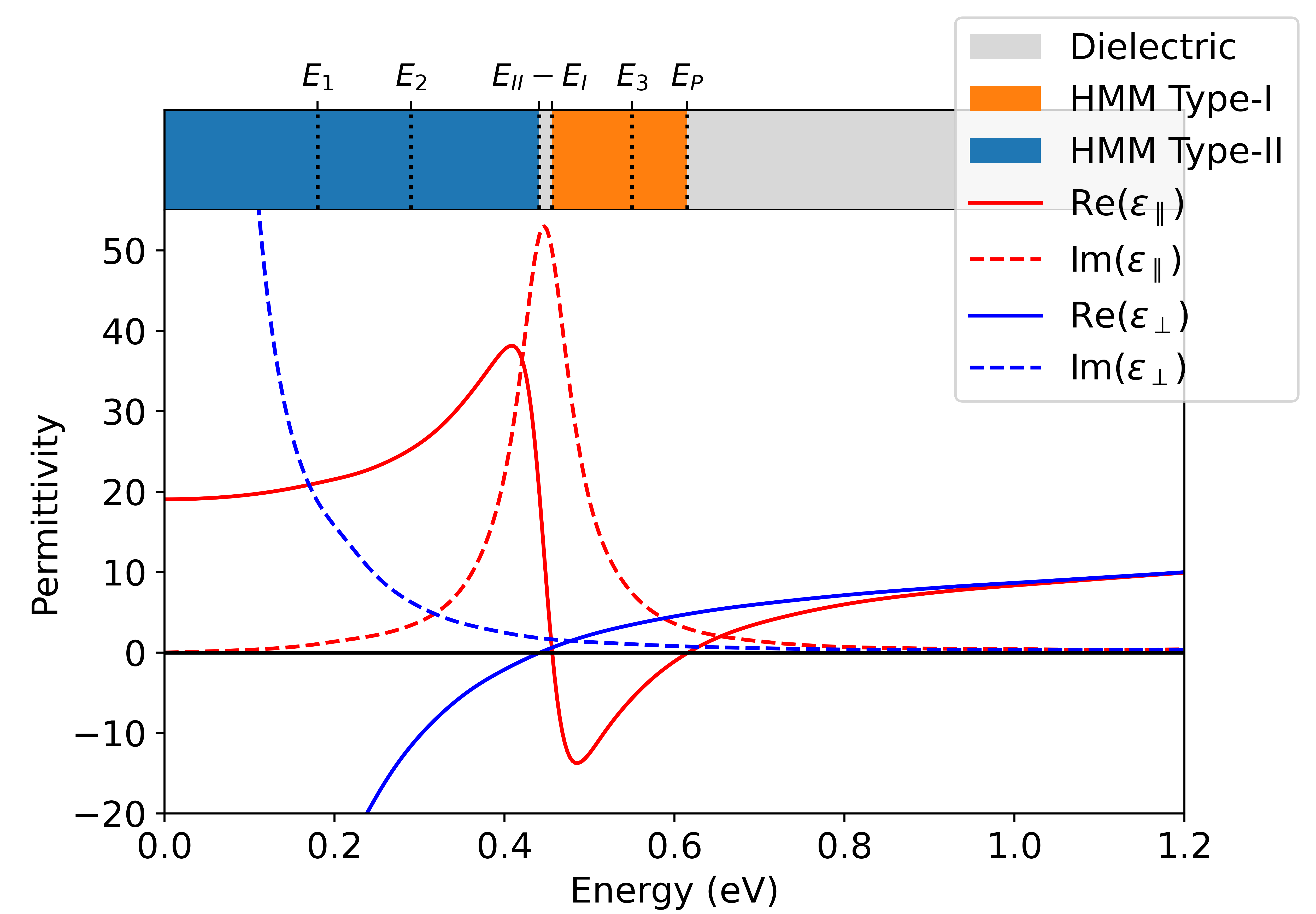}
\caption{\small Real (solid lines) and imaginary (dashed lines) parts of the
parallel (red lines) and perpendicular (blue lines) dielectric functions of
Si:InAs/AlSb multilayer
($c_{\bf 2}=3.12\%$ and $f_m$ = 0.5). Top bar
indicates spectral optical character of the multilayer -- namely Type-I (orange) and Type-II (blue) hyperbolic, and dielectric (gray). Vertical dashed lines mark characteristics energies discussed in the text.}
\label{fig:fig3}
\end{figure}

\noindent{\bf Hyperbolic multilayers.}
By treating metamaterials as homogeneous, EMT predicts the hyperbolic behavior of the multilayers and the spectral features of the VPPs within the stacks. 
The optical properties of HMMs are evaluated by combining the bulk dielectric functions of the intrinsic III-V semiconductors ($\epsilon_d$)
 and Si:InAs ($\epsilon_m$) calculated from first principles. Equations (\ref{eps_perp}-\ref{eps_par}) 
are used to compute the perpendicular ($\eperp$) and parallel ($\epar$) components of the effective dielectric tensor with respect to the
optical axis (Figure \ref{fig:fig1}a). 
Nine multilayers have been considered by mixing the III-V semiconductors with model {\bf 2} of Si:InAs. 
Figure \ref{fig:fig3} shows the real (solid lines) and imaginary (dashed lines) parts of the
parallel (red lines) and perpendicular (blue lines) dielectric functions of equal-composition ($f_m=f_d = 0.5$) Si:InAs/AlSb stack, taken as reference. The results for the remaining  Si:InAs/III-V multilayers are summarized in Section S4 of SI.

\begin{figure*} [!t]
\centering
\includegraphics[width=0.9\textwidth]{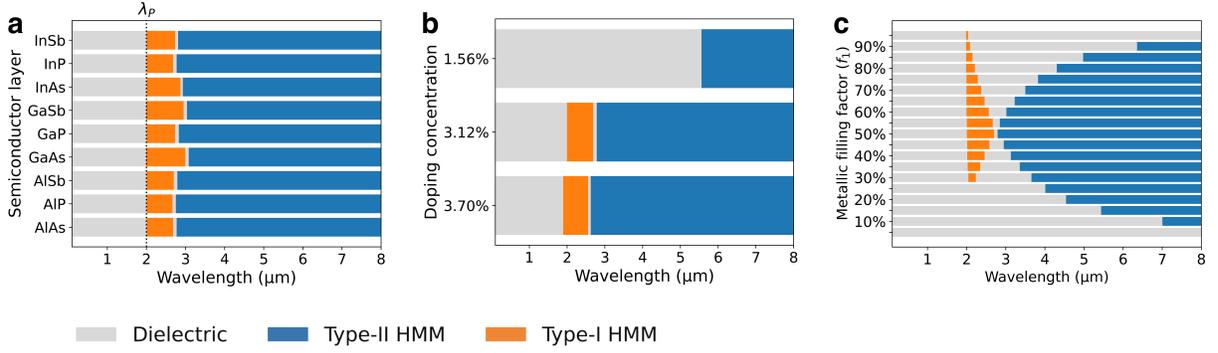}\\
\caption{Spectral optical character -- namely Type-I (orange) and Type-II (blue) hyperbolic, and dielectric (gray) --  of 
    HMMs composed of: a) III-V semiconductors/Si:InAs ($c_{\bf 2}=3.12\%$, $f_m=0.5$)
    as a function of the dielectric component; b) Si:InAs/AlSb ($f_m=0.5$) as a function of Si doping concentration; c)  Si:InAs/AlSb ($c_{\bf 2}=3.12\%$)
    as a function of the filling factor $f_m$  in the range (0.05 -- 0.95)\%.}
\label{fig:fig4}
\end{figure*}

The hyperbolic dispersion implies the condition $\eperp\cdot\epar< 0$.
At low energy ($E<E_{II}=0.44$ eV in Figure \ref{fig:fig3}), Re[$\eperp$] is dominated by
$\epsilon_m$ and Re[$\epar$] by $\epsilon_d$. This implies that the perpendicular component (blue lines) has
a metal-like behavior similar to Si:InAs, while the parallel component (red
lines) has a typical dielectric character similar to those of the undoped
semiconductor. Thus, the multilayer acts as Type-II HMM. 
In the range $E\in\{0.44-0.47\}$ eV, Re[$\eperp$] has a
zero, switching from negative to positive for all systems.  This happens when Re[$\epsilon_m$]$=-$Re[$\epsilon_d$], which in turn
corresponds to a maximum of $\epar$. Beyond this resonance energy,  Re[$\epar$] decreases until it crosses the zero, switching sign from positive to negative, 
with a typical Lorentz-like character. The energy value of the Lorentz resonance depends on the III-V semiconductor
used as dielectric. For $E>E_I=0.47$ eV the trend is inverted, 
with $\eperp$ driven by $\epsilon_d$ and $\epar$ by $\epsilon_m$. As a consequence,  
$\epar$ has a metal character, while $\eperp$ has a dielectric
behavior, corresponding to a Type-I HMM. 
For $E > E_p=0.62$ eV  both components (III-V and Si:InAs) are dielectric-like and the
same holds for the multilayer.

A direct comparison with the experimental results of Refs. \cite{sohr2018,sohr2021} indicates that our theoretical spectra are in very good agreement with all the experimental features of the multilayer, including the
alternation of the hyperbolic type character as a function of the radiation energy. For a better comparison with experimental data, Figure S11 of SI reports the dielectric function of Figure \ref{fig:fig3}) as a function of wavelength, instead of energy.
Yet, our theoretical spectra are affected by a slight blue shift, when compared with the experiments.
This minor discrepancy is mostly ascribable to: (i) the residual band gap
underestimation of the pristine semiconductors calculated at the DFT+$U$
level; (ii) the exact doping level, which may differ from
the exact experimental value; (iii) the assumption of ideal, defectless interfaces in the theoretical modeling.
Despite these minor details, the agreement with the experimental results
confirms {\em a posteriori} the accuracy of the joint DFT+$U$/EMT approach
used in this work.

The effects of composition, thickness, and doping on the hyperbolic behavior of the multilayers are illustrated in Figure \ref{fig:fig4}. 
To better appreciate the differences in the optical properties of HMMs in the low energy part of the spectrum, these results are expressed in terms of 
wavelength ($\lambda$) instead of  energy.
The nine metamaterials (Figure \ref{fig:fig4}a) exhibit a variable  behavior as a function of the
radiation wavelength. The hyperbolic character appears for $\lambda > \lambda_p = {hc}/{E_p}= 2.0~\mu$m (i.e., $E < E_p =0.62$ eV). 
All systems have a prevalent Type-II character (blue area), except for an intermediate range,
where they exhibit a Type-I dispersion (orange area). The former type
 is correlated to the negative value of $\eperp$, while the latter is associated with the  Lorentz-like
inversion of $\epar<0$. At shorter wavelengths ($\lambda < \lambda_p$), the radiation overcomes the plasma frequency of the metal and the multilayer is
globally dielectric (gray area). 
Since the plasmon energy of Si:InAs varies with the doping (Table \ref{tab:tab1}), the hyperbolic range can be 
tuned by controlling the dopant concentration. An increase in the doping concentration (Figure \ref{fig:fig4}b)
 imparts a blue shift of the plasma frequency that sets the transition between the hyperbolic and the elliptic (i.e., dielectric) dispersion behavior. In the case of $c_{\bf 1}=1.56\%$, and $f_m = 0.5$, the Si:InAs/AlSb multilayer exhibits only a Type-II character in the hyperbolic range. However, for metallic filling factors between 0.70 and 0.95, the system also develops narrow frequency intervals in which it displays a Type-I hyperbolic response. Supplementary plots (Fig. S12, SI) -- analogous to Figure \ref{fig:fig4}c -- for the lowest ($c_{\bf 1}$) and highest ($c_{\bf 3}$) doping concentrations confirm that the onset of Type-I and Type-II hyperbolicity depends on both doping level and metallic filling factor.

Figure \ref{fig:fig4}c displays  the trend associated with the variation of the composition filling factor $f_m$, which measures the amount of the metallic component within the
multilayer: $f_m = 0$ ($f_m = 1$) corresponds to pure dielectric (metallic) system.
For $\lambda > \lambda_p$, the actual hyperbolic
range depends on the filling factor. 
In general, a metal/dielectric ratio of about $40-60\%$ provides a wider spectral range with hyperbolic dispersion,
which progressively reduces to longer wavelengths as much as one of the components becomes predominant.
In summary, by controlling the chemical species, the doping, and the geometry of the stack it is possible to fine tune the optical
character of the multilayers and achieve hyperbolic behavior across the near-IR to THz spectral range.

\noindent{\bf Volume plasmon-polaritons.}
The spectral analysis of VPP excitations is obtained from the angular dielectric function $\epsilon_{\varphi}(E)$  and the propagation angle $\Theta(\varphi)$, where $E$ is the energy and 
$\varphi$ is the angle between the wavevector  of the incoming radiation  {\bf k} within the metamaterials and the optical axis {\em z} (see Sec. S1 of SI, for further details).
Here, we focus on the reference case
of Si:InAs/AlSb multilayer (doping concentration $c_{\bf 2}=3.12\%$ and filling factor $f_m=0.5$).
Since $\epsilon_{\varphi}$ is a function of the energy of the incoming radiation, we selected 
a few representative energies ($E_i$) marked by vertical dashed lines in Figure \ref{fig:fig3}.
Energies $E_1=0.18$ eV and $E_2=0.29$ eV correspond to a Type-II hyperbolic character, while $E_3=0.55$ eV corresponds to  a Type-I HMM. 

\begin{figure}[!t]
\centering
\includegraphics[width=0.45\textwidth]{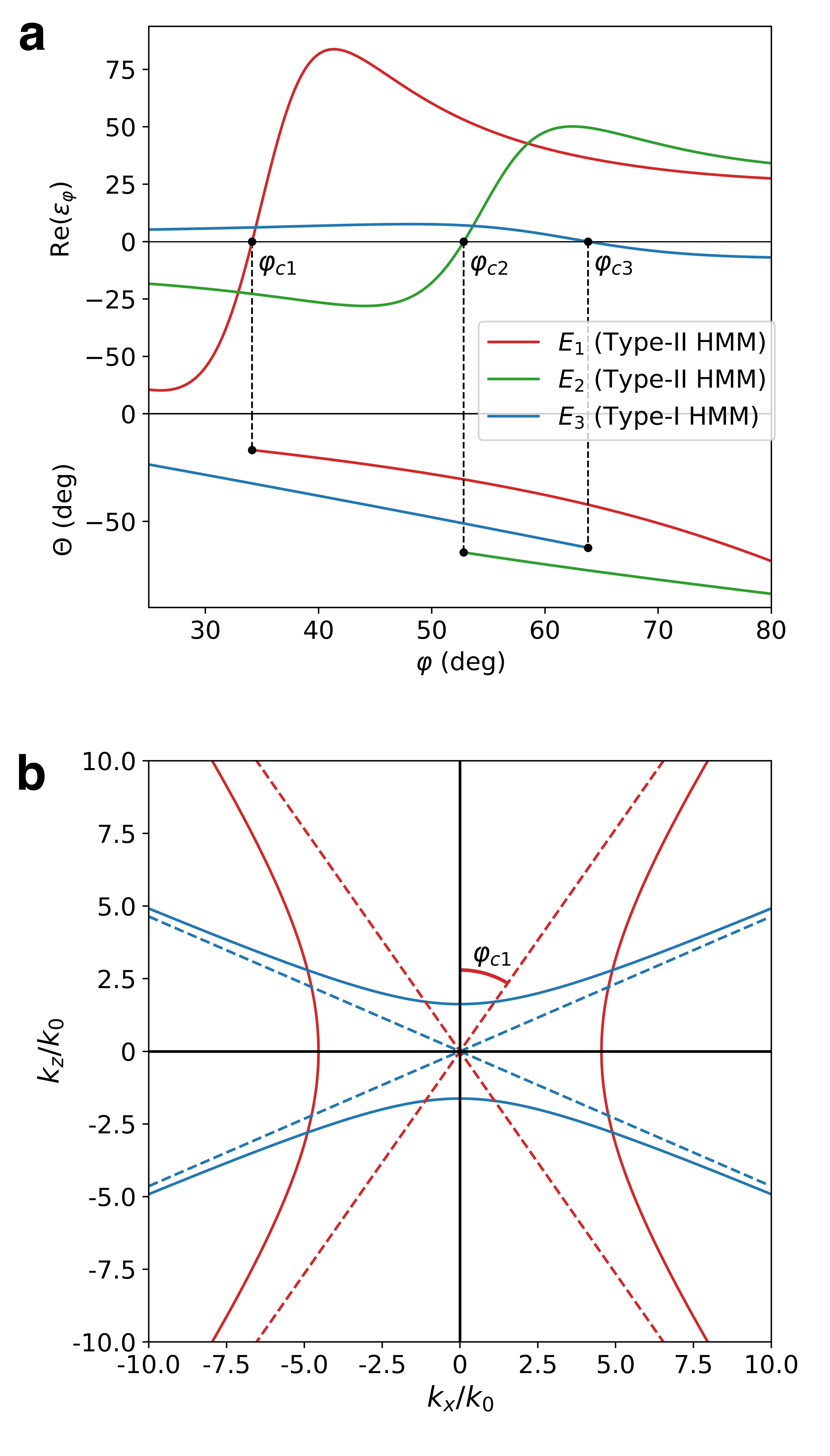}
\caption{\small a) Real part of the angular dielectric function ($\epsilon_\varphi$, top panel) and allowed conic angles ($\Theta$, bottom panel) of excited VPPs with respect to optical axis  for Si:InAs/AlSb multilayer ($c_{\bf 2}=3.12\%$, $f_m=0.5$) corresponding to the energies $E_1$, $E_2$, and $E_3$ shown in Figure \ref{fig:fig3}; b) 2D projected $k$-dispersions for Si:InAs/AlSb (straight lines) corresponding to $E_1$ (Type-I, red lines) and $E_2$ (Type-II, blue lines). Dashed lines correspond to hyperboloid asymptotes whose slopes correspond to critical angles  $\varphi_{c}$ marked in panel a.}
\label{fig:fig5}
\end{figure}

In the case of $E_1$ and $E_2$, the real part of $\epsilon_{\varphi}$ is negative at small angles and it  becomes positive beyond a critical angle $\varphi_c$, which increases when the energy is increased ($\varphi_{c1}<\varphi_{c2}$). 
The trend is inverted for $E_3$  where $\epsilon_{\varphi}$ is positive at small angles and changes sign at $\varphi_{c3} \sim 65^{\circ}$. 
This behavior follows the distinct topology of the underlying hyperboloids that describe the dispersion isosurfaces --two-sheeted in the Type-I case and one-sheeted in the Type-II case. Indeed,
the condition $\mathrm{Re}[\epsilon_{\varphi}]=0$ determines the angular boundary between the metallic and dielectric response of the metamaterial (the analogous of plasmon excitation in bulk materials),  and determines  the critical angle $\varphi_c$  that satisfies the wavevector refraction conditions at the interface between the metamaterial and the external environment (Figure \ref{fig:fig1}a). For each energy, $\tan(\varphi_c)=k_z/k_x=\sqrt{\eperp/\epar}$ represents the asymptote to the hyperboloid that describes the dispersion isosurface of the metamaterial (Figure \ref{fig:fig5}b). Thus, for a Type-I (Type-II) HMM, $\varphi_c$ is the maximum (minimum) angle that 
the incoming radiation must have to propagate within the stack. 
As such, the three intersections $\varphi_{ci}$ in Figure \ref{fig:fig5}a correspond to the extreme geometrical conditions  for the excitation of conic VPPs, whose angular radii are given by the corresponding $\Theta$ angles (Figure \ref{fig:fig5}a, bottom panel).
For all the considered energies,  $\Theta$ is negative for all the allowed $\varphi$ angles. This is the signature of the excitation of backward waves, where the Poynting vector {\bf S} and the wavevector {\bf k} are in opposite, lateral
directions. This anomalous refraction effect is the basis for advanced
transformation optics and EM cloaking applications.

In real experimental conditions, the angle $\varphi$ is not a free variable but  is  dictated by the number of bilayers in the stack as well as by the matching setups (e.g., prism or grating) used to compensate the momentum mismatch between the incident radiation and the VPP. This means that, given the actual geometry of the overall system (HMM $+$ coupler), only a few energies and specific angles $\varphi$ may effectively excite a VPP. These aspects, along with the multiscattering effects deriving from the internal metal/dielectric interfaces, cannot be caught by EMT that considers the entire multilayer as a single homogenized medium.  A proper description of these effects is necessary to be predictive in designing/optimizing customized heterostructures.

In order to overcome these limitations, we solved the electromagnetic problem of a finite multilayer composed of alternating metal/dielectric materials, by employing the scattering matrix method. This approach allows us to model the behavior of the EM waves that propagate through the metamaterial and to explicitly include the boundary conditions imposed by the grating layer.
Along the line of the experimental samples \cite{sohr2018,sohr2021}, we considered a metamaterial composed of $\mathcal{N}=10$ periods of alternating Si:InAs/AlSb bilayers, with the following parameters:
doping concentration $c_{\bf 2}=3.12\%$, filling factor $f_m=0.5$, and thickness of a single bilayer $d_{tot}=100$ nm (Figure \ref{fig:fig1}a). The dielectric functions that characterize the individual layers are those calculated at the DFT+$U$ level for the respective bulk materials. A gold grating layer with variable periodicities $\Lambda =$ 0.9$~\mu$m, 1.3 $\mu$m, and 1.8 $\mu$m, respectively, is introduced on top of the multilayer stack. The value $\Lambda =$ 1.8 $\mu$m
corresponds to the experimental condition used in Ref. \cite{sohr2018}. The entire system -- comprising both multilayer and grating -- is embedded in air, modeled as a medium with dielectric constant $\epsilon=1.0$.

\begin{figure} [!t]
\centering
\includegraphics[width=0.4\textwidth]{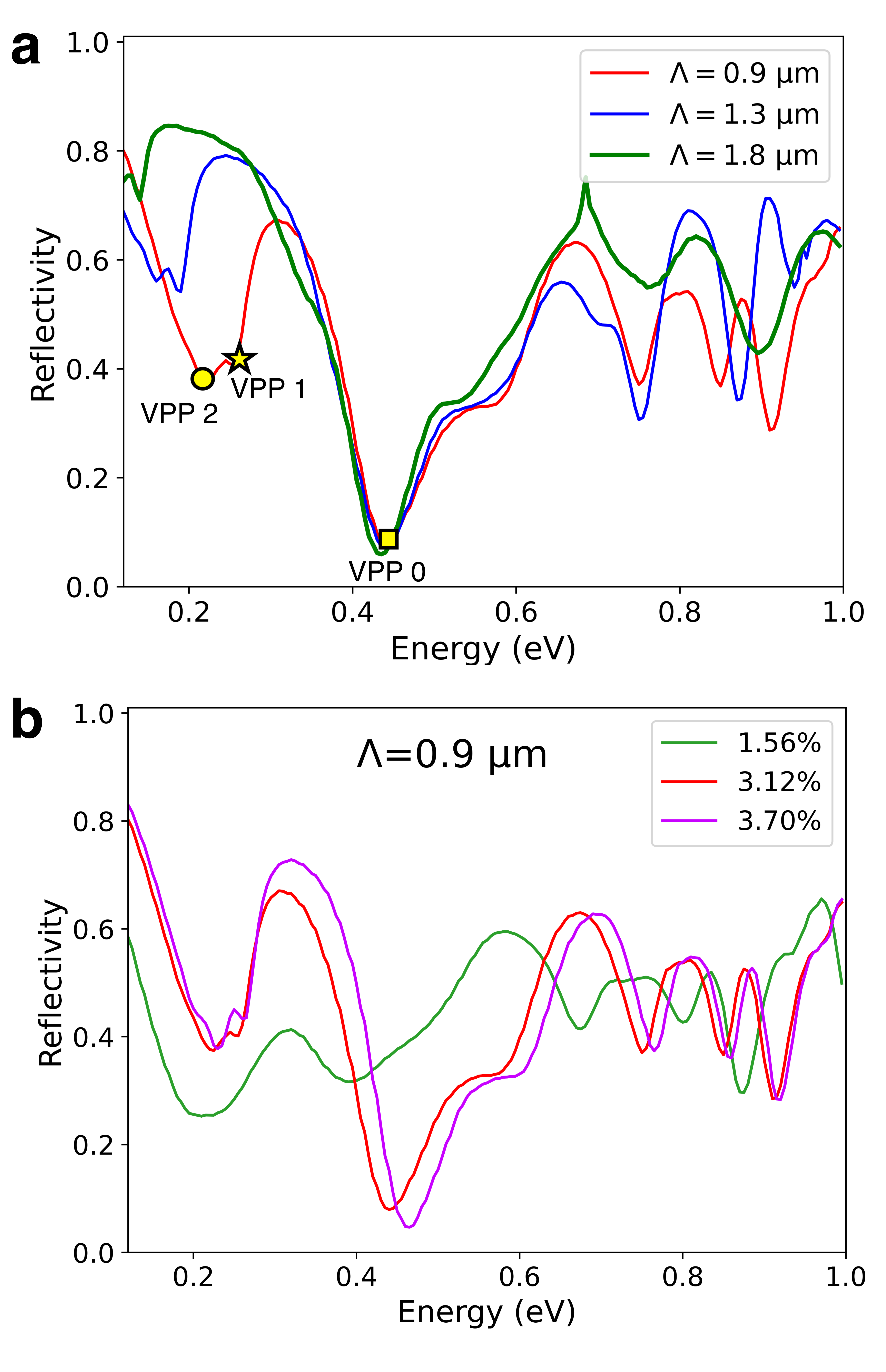}\\
\caption{S-matrix simulations of reflectivity  spectra (R) for Si:InAs/AlSb multilayers as a function of: a)  grating periodicity ($\Lambda$);  b) Si-doping concentration. Yellow square, star and circle correspond to VPP0, VPP1, and VPP2, respectively.}
\label{fig:fig6}
\end{figure}

Figure \ref{fig:fig6}a shows the  reflectivity spectra (R) calculated by varying the spacing of the grating coupler. 
In the hyperbolic energy range ($E<E_p=0.62$ eV), the excitation of VPP modes corresponds to local minima in the reflectivity spectra. For example, in the case of $\Lambda =$ 0.9 $\mu$m, three distinct resonances -- labeled VPP0, VPP1, and VPP2 and marked with yellow symbols -- can be identified. The corresponding transmittance and absorption spectra are reported in Figure S13 of the SI. Resonance VPP0 occurs in the Type-I hyperbolic region, while VPP1 and VPP2 occur in the Type-II domain. Here, the hyperbolic character is derived from EMT (Figure \ref{fig:fig3}a). The remaining minima for $E>E_p$ correspond to usual interband excitations in dielectrics. VPP0 has the most pronounced reflection minimum, hardly modified by the grating geometry. The increase in the grating period causes a reduction of the reflection dips. 
 Even though these three VPP modes lie in the mid-IR, this does not limit the operational range of the HMM to the THz region. Indeed, these VPP energy values result from a specific combination of doping, stack geometry, and  momentum-matching conditions imposed by the grating coupler.  When the grating period is increased (especially for $\Lambda=1.8 \mu$m), the VPP branches  associated with  VPP1 and VPP2 undergo a red-shift with respect to the $\Lambda=0.9 \mu$m case. These red-shifted modes fall squarely in the mid-IR to THz range, in agreement with the experimental findings \cite{sohr2018,sohr2021}.

\begin{figure*} [!t]
\centering
\includegraphics[width=0.9\textwidth]{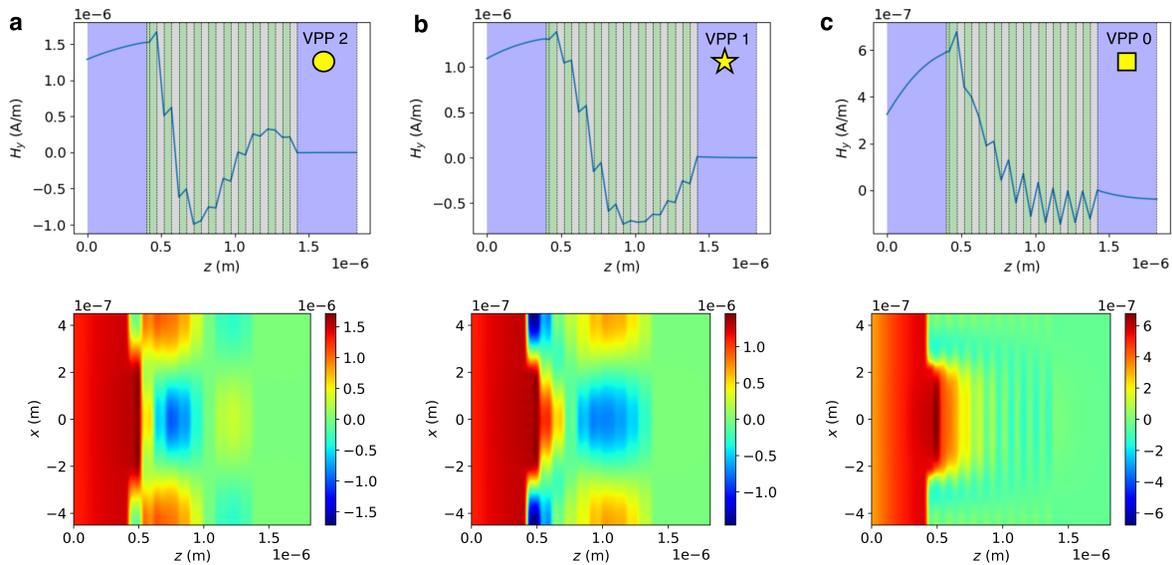}\\
\caption{Magnetic field intensity (top) and 2D profile (bottom) of the  a) VPP2, b) VPP1, and c) VPP0 modes identified in Figure \ref{fig:fig6}a  for  Si:InAs/AlSb HMM ($c_{\bf 2}=3.12\%$, $f_m=0.5$, $d_{tot}=100$ nm, $\mathcal{N}=10$, and  $\Lambda =0.9 \mu$m). Green/gray areas identify metal/dielectric bilayers, violet areas mark the reflection/transmission embedding media (i.e., air); incoming light impinges on the left side of the multilayer, which includes also the grating layer. The 2D field profiles are computed with a spatial resolution of $\sim $11 nm along the {\em x} direction and $\sim$23 nm along the {\em z} direction.}
\label{fig:fig7}
\end{figure*}

For non magnetic HMMs, only transverse magnetic modes can be excited \cite{Lindell:2001cu}. The magnetic fields profile corresponding to the three VPPs marked in Figure \ref{fig:fig6}a are shown in Figure \ref{fig:fig7} (upper panel), which actually confirm that the EM waves propagate within the stack.  However, only VPP0  wave propagates  with significant intensity outside the multilayer, i.e., $H>0$ in the region on the right. For VPP1 and VPP2, the front-grating allows for the excitation of the EM modes within the multilayer, while
the low-$k$ mismatch -- typical of Type-II HMMs --  prevents their transmission into the air region (see also the transmittance plot in Figure S13 of  SI). As a result, these modes can be excited but remain confined within the stack. A different external dielectric material or the inclusion of a proper back-grating could facilitate the propagation of these modes outside the multilayer.

The VPP modes can be classified into distinct orders, based on the number of nodes in the magnetic field within the metamaterial (Figure \ref{fig:fig7}, lower panel). 
VPP0 has zero nodes and represents the {\em 0}-th order mode; VPP1 and VPP2 have one and two nodes and correspond to the {\em 1}-st and {\em 2}-nd order modes, respectively. 
The modulation of doping in the conductive layers further impacts the VPP modes (Figure \ref{fig:fig6}b), by varying the hyperbolic ranges as well as the number and the energy position of the minima in the reflectivity.  In particular, low-dopant concentration causes an overall flattening of the reflection spectrum, which indicates a higher energy loss of VPP modes.
This implies that it exists a minimum doping level to effectively sustain plasmon-polariton waves across the multilayer.

\begin{figure}[!t]
\centering
\includegraphics[width=0.5\textwidth]{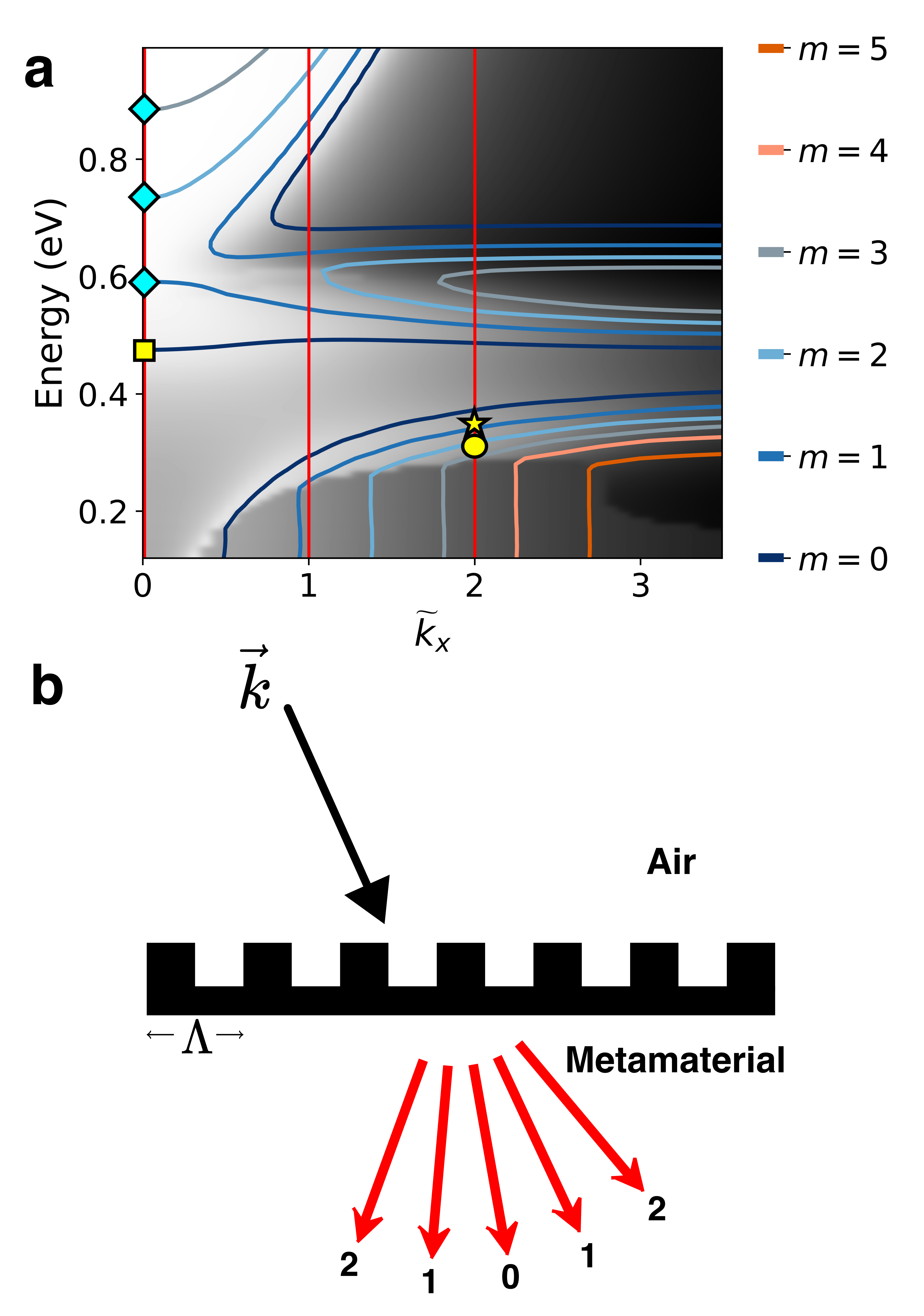}
\caption{\small a) Photonic band structure of Si:InAs/AlSb ($c_{\bf 2}=3.12\%$, $f_m=0.5$, $d_{tot}=100$ nm). Red vertical lines correspond to lowest harmonic contributions for $\Lambda=0.9$ $\mu$m, colored curved lines correspond to the first six $k_z$  resonance values ($m=\{0-5\}$) for $\mathcal{N}=10$ metal/dielectric bilayers. Colored symbols indicate EM  solutions for the selected geometric conditions: cyan diamonds correspond to ordinary dielectric refracted modes,  yellow  symbols indicate the most intense VPP modes as in Figure \ref{fig:fig6}a. b) Schematic representation of harmonic generation from planar grating. Integer labels indicate the lowest diffracted harmonics and correspond to red lines in panel a). 
 } 
\label{fig:fig8}
\end{figure}

This analysis confirms that this class of metamaterials may sustain hyperbolic dispersion and VPP propagation down to THz frequencies, while the details of the hyperbolic response critically depend on the number of layers and the grating geometry. For this reason, it is useful to calculate the photonic band structure of the periodically repeated multilayer, which provides the overall distribution of electromagnetic  modes  of the system (i.e., Bloch states). The results for Si:InAs/AlSb ($c_{\bf 2}=3.12\%$, $f_m=0.5$, $d_{tot}=100$ nm) are summarized in Figure \ref{fig:fig8}.
Because of continuity conditions of the EM fields at the interface, the energy $E$ and the wavefunction components parallel ($k_x$) and  perpendicular ($k_z$) to the interface (Figure \ref{fig:fig1}a) are not independent variables.
In Figure \ref{fig:fig8}a,  the pixel colors in the grayscale spectrum represent the magnitude of the imaginary part of the complex Bloch wavevector $k_z$ as a function of $E$ and $k_x$. Lighter regions correspond to areas where the imaginary component is smaller, indicating less attenuation of electromagnetic waves within the metamaterial. Conversely, darker regions indicate higher attenuation due to a higher imaginary component. 

While the PBS plot provides continuous solutions for the excitation of EM modes within the infinite multilayer, the application of boundary conditions due to finite-thickness stack and matching coupler imposes selective rules, which restrict the possible VPP modes that can be excited in the metamaterial. For example, the inclusion of a grating layer imposes conditions on the in-plane $k_x$ wavevectors, which depend on the grating periodicity $\Lambda$. For this reason,
in  Figure \ref{fig:fig8}a we plotted the bandplot as a function of $\tilde{k}_x=k_x/T_x$, where $T_x=2\pi/\Lambda$, and $\Lambda=0.9 \mu$m. This allows for the decomposition of grating refraction components in multiple harmonics as shown in  Figure \ref{fig:fig8}b. The integer values $\tilde{k}_x=0, 1,\ldots,n$ (vertical red lines) correspond to the fundamental, the first, $\ldots$, the $n^{th}$ harmonic, respectively; the higher the harmonic, the lower  the intensity of the refracted EM component. 

In a similar way, the finite number of repeated metal/dielectric bilayers selects specific values for the possible $k_z$ perpendicular to the stack. It is possible to excite just a finite number of Bloch waves in the multilayer, and this happens when the thickness of the stack is an integer multiple of half a wavelength of the incoming radiation. This corresponds to satisfy the resonance condition
${d_{tot}}/2\pi k_z=({m+1})/{2 \mathcal{N}}$,
where $d_{tot}$ and $\mathcal{N}$ are the thickness and the number of the repeated bilayers, respectively, and $m$ is an integer ranging from $0$ to $\mathcal{N}-1$. The $m$-index indicates the number of nodes of the EM field within the metamaterial and thus provides a direct correlation between the resonance values of $k_z$ and the order of the VPP discussed above. Colored lines in Figure \ref{fig:fig8}a represent the isovalue lines corresponding to the first six resonance $k_z$ values for a Si:InAs/AlSb composed of $\mathcal{N}=10$ bilayers. 
Once the grating and the thickness conditions are fixed, the intersections between the vertical red lines and the colored isolines correspond to the unique possible EM fields that may be excited within the metamaterials. Changing the values of $\Lambda$ and/or $\mathcal{N}$ shifts the position of the matching points in the photonic band structure plot. 

In the case of Figure \ref{fig:fig8}a, for $E>0.6$ eV the system has a  dielectric character, thus the resonance points (cyan diamonds) correspond to the ordinary light refraction conditions.  The values E=0.74 eV and 0.89 eV at $\tilde{k}_x=0$ closely match the higher energy minima in the reflectivity spectrum of Figure  \ref{fig:fig6}a. 
 For $E\sim0.5$ eV, the system has a Type-I hyperbolic response, and the matching point between the fundamental harmonic line $\tilde{k}_x=0$ and the $m=0$ isoline represents the VPP0 mode (yellow square), which corresponds to the lowest minimum in the reflectivity spectrum. The flat energy dispersion of the $m=0$ isoline concurs with the insensitivity of the reflection minimum with respect to the grating periodicity $\Lambda$ shown in Figure  \ref{fig:fig6}a. 
At lower energy, the system has a Type-II character. In this case, the fundamental harmonic does not contribute to higher order VPPs, i.e., no isoline intersections at  $\tilde{k}_x=0$. The VPP1 (yellow star) and VPP2 (yellow circle) resonances observed in reflection spectra mostly stem from the second harmonic diffraction peak ($\tilde{k}_x=2$) and the isoline curves for $m=1$ and $m=2$. The proximity of these peaks to dark-color edge in the plot (i.e., higher energy dissipation) goes in parallel with the lower intensity of the corresponding reflection minima. Higher order VPPs could be further excited at higher harmonic conditions, but their low intensities make them hardly recognizable in the reflectivity spectra.  

\section*{Conclusion}
By combining multiscale and multiphysics techniques, we have investigated the possibility to excite and tune volume plasmon-polaritons modes in hyperbolic metamaterials based on III-V semiconductors. We analyzed the multi-dimensional phase space, by varying chemical composition, doping concentration, metallic filling factor, number of layers in the stack, and the grating coupler. This allows us to have a direct comparison with experiments and to gain a microscopic understanding of the origin of the volume plasmon-polaritons modes in these hyperbolic metamaterials. First principles simulations are used to evaluate the effect of doping on the dielectric function and the plasmonic excitation of the composing materials. While  effective medium theory  provides insights on the overall optical response of the metamaterials, the $S$-matrix approach is shown to be a powerful tool for interpreting  experimental spectra, which keeps directly into account all the geometric features of the experimental setups. Finally, the photonic band structure is particularly useful to design and optimize the geometrical conditions to tune the energy and the spatial distribution of the volume plasmon-polaritons that may travel across the multilayer. 
Our integrated approach paves the way to a unique material-geometry co-design to realize optimal plasmon-based hyperbolic metamaterials. Starting from an atomistic description of the constituent materials, we demonstrate that the proper choice of composition and geometry setup allows for the excitation of multiple VPP traveling modes with high quality factors (i.e., low losses) across the mid-IR and THz ranges. Their energies and propagation lengths can be selectively controlled, opening new routes for bio-sensing applications, thermal emission control, and THz detection. In particular, the choice of III-V semiconductors (rather than noble metals) promises better compatibility with epitaxial growth, improving the monolithically integration with other semiconductor nanostructures (e.g., 2D materials, quantum wells, quantum dots, etc.), and the realization of novel hybrid polaritonic THz devices (e.g., on-chip optics, lasers, detectors, modulators) with enhanced performances and novel dispersion engineering. Furthermore, the high photonic density of states of HMMs along with the plasmon tunability of doped-semiconductors make these systems promising for enhanced thermal emission and radiative thermal transfer, also beyond black-body limit.
    
    \section*{Acknowledgements}
L.B. acknowledges financial support from PNRR MUR project ECS\_00000033\_ECOSISTER. 
A.C. acknowledges the National Centre for HPC, Big Data and Quantum Computing (ICSC), funded under the National Recovery and Resilience Plan (NRRP), Mission 04 Component 2 Investment 1.4, NextGenerationEU, Award Number:CN00000013.
S.Cur. acknowledges support by the Office of Naval Research under grants N00014-23-1-2615 and N00014-24-1-2768, and by the DoD High Performance Computing Modernization Program (Frontier).
S.Cur. also thanks Auro Scientific, LLC for computational support.
The authors thank Drs. Simon Divilov, Xiomara Campilongo and Doug E. Wolfe for useful discussions.
 
     \section*{Supporting Information}
    Supplementary Information includes a methodological complement of the EMT, SMM and PBS approaches introduced in Section 2 and used in this work. The section of SI collects a compilation of additional results: the structural and electronic properties of III-V semiconductors;  optical properties of III-V semiconductors, including the frequency-dependent dielectric functions;  EMT characterization of HMMs composed of III-V semiconductors/Si:InAs ($c_{\bf 2}=3.12\%$, $f_m=0.5$) as a function of the dielectric constituent, as well as the hyperbolic character of AlSb/Si:InAs HMM as a function of both doping concentration and metallic filling factor;  full reflectivity, transmittance, and absorption spectra for finite AlSb/Si:InAs HMM ($c_{\bf 2}=3.12\%$, $f_m=0.5$, $\mathcal{N}=10$) obtained from SMM simulations.
       
    \section*{Author Contributions} 
A.C. and L.B. conceptualized the work. S.C. performed calculations and implemented the AFLOW-EMERALD code. 
S.C., L.B., and A.C. analyzed data. A.C. and S. Cur. established the research
direction and supervised the project. All the authors contributed to the writing of the manuscript.

    \section*{Notes}
    The authors declare no competing financial interest.

\bibliography{bibliography_Rev}
 \clearpage
 
 \section*{SUPPORTING INFORMATION} 
\renewcommand{\thefigure}{S\arabic{figure}}
\renewcommand{\thetable}{S\arabic{table}}
\setcounter{figure}{0}
\section*{Methodology}
\begin{figure}[b!]
	\centering
	\includegraphics[width=0.85\linewidth]{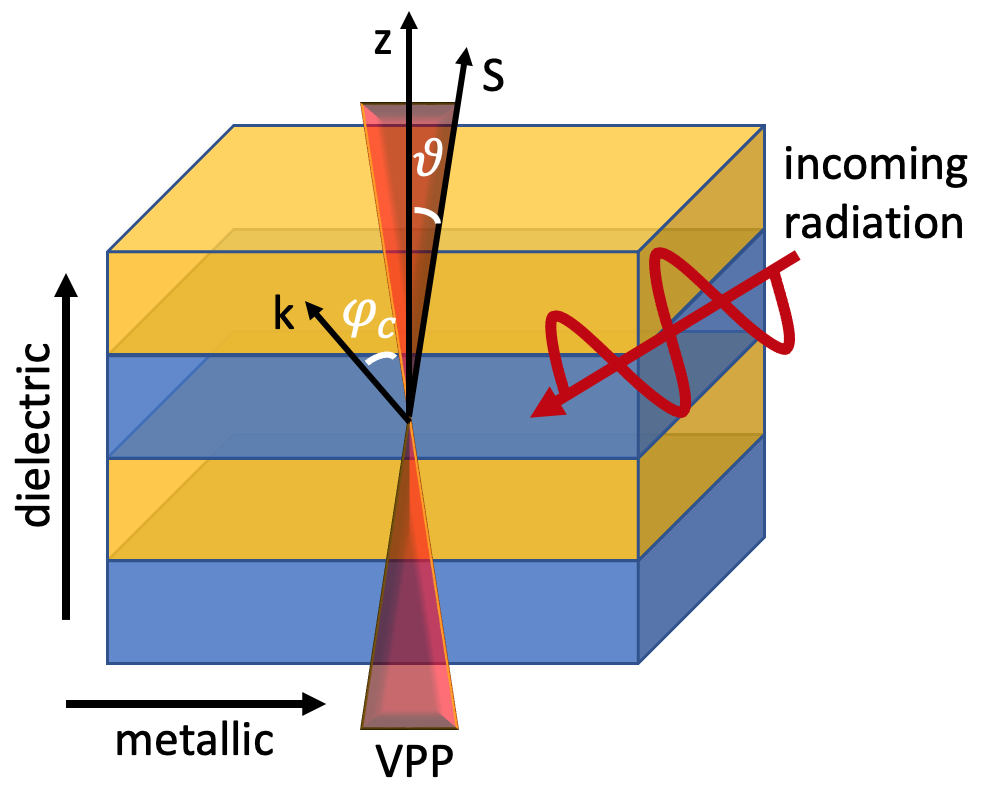}
	\caption{Schematic vector diagram for transverse magnetic (TM) propagating waves and hyperbolic dispersion isosurface corresponding to a type-II HMM.}
	\label{cone}
\end{figure}

\noindent {\bf Volume-plasmom polaritons from EMT}. 
The macroscopic conditions for  VPP excitation are derived from the angular dielectric function \cite{Ishii:2013er}:
\begin{equation}
\epsilon_{\varphi}(E)=\frac{\eperp\epar}{\eperp \sin^2(\varphi)+\epar \cos^2 (\varphi)},
\label{eq:phi}
\end{equation}
where $E$ is the energy and $\varphi$ is the angle between the wavevector  of the incoming radiation  {\bf k} within the metamaterials and the optical axis {\em z}, as schematized in Figure \ref{cone}; 
$\epar$ and  $\eperp$ are effective dielectric functions in the directions parallel  and perpendicular  to the optical  axis, as defined in Eqs. 1-2 (Section 2) and in Figure 1a of main text. 

The angle $\Theta$ between the extraordinary VPP wave (i.e., the Poynting vector) and the optical axis is:
\begin{equation}
\Theta(\varphi)=\tan^{-1} \bigg(\mathrm{Re}\bigg[\frac{\epsilon_{\parallel}} {\epsilon_{\perp}}\bigg] \tan(\varphi) \bigg).
\label{eqtheta}
\end{equation}

\noindent {\bf Scattering Matrix Method}.  
The matrix $\mathbf{S}^{(n)}$ relates the amplitude coefficients of the forward ($c^+$) and backward ($c^-$) propagating components of the electric and magnetic fields at the left (1) and right (2) interfaces of the {\em n}-th layer, as illustrated in Figure 1b of main text. Combining $\mathbf{S}^{(n)}$ of all layers, one obtains the total scattering matrix of the multilayer  in free space
$\mathbf{S}^{(\text {ML)}}= \mathbf{S}^{(1)} \otimes \mathbf{S}^{(2)} \otimes \cdots \otimes \mathbf{S}^{(\mathcal{N})}$.
The multilayer properties can be easily connected to the external environment by introducing
the matrices $\mathbf{S}^{(\text {r})}$ and $\mathbf{S}^{(\text {t})}$ that account for the reflection and  transmission embedding regions, i.e., the media from which the incident light originates and into which the transmitted light propagates, respectively. 
The total scattering matrix of the system is: $\mathbf{S}^{(\text {tot)}}=\mathbf{S}^{(\text {r})} \otimes \mathbf{S}^{(\text {ML})} \otimes \mathbf{S}^{(\text {t})}$.
From $\mathbf{S}^{(\text {tot)}}$ the mode coefficients of the reflected and transmitted EM field can be obtained as:
$c_r=S_{11}^{(\text {tot)}}c_{in}$ and 
$c_t=S_{21}^{(\text {tot)}}c_{in}$,
where $c_{in}$, $c_{r}$, $c_{t}$  are the amplitude coefficients of the incoming, reflected and transmitted radiation, respectively. The reflectivity (R), transmissivity (T), and absorption (A) functions are calculated from the resulting EM fields being reflected, transmitted, and absorbed  through the entire multilayer \cite{emerald}.

\noindent {\bf Photonic Bandstructure}.  
Photonic bandstructures are calculated with the {\em AFLOW-EMERALD} code \cite{emerald} by expanding  the electric and magnetic fields on a plane waves basis set, 
which satisfy both the Bloch theorem and the divergence equation.  In the case of magnetic field, the real space Maxwell's equation:
\begin{equation}
\nabla \times \Big[\frac{1}{\epsilon(\mathbf{r})}\nabla \times \mathbf{H}(\mathbf{r})\Big]=\frac{\omega^2}{c^2}\mathbf{H}(\mathbf{r})
\end{equation}
is transformed into a matrix relation: 
\begin{equation}
\sum_{G'}\mathbf{H}_{G,G'}c_n(\mathbf{k}+\mathbf{G'})=\frac{\omega^2}{c^2}c_n(\mathbf{k}+\mathbf{G}),
\end{equation}
where 
\begin{equation}
\mathbf{H}_{G,G'}=\vert\mathbf{k}+\mathbf{G}\vert\vert\mathbf{k}+\mathbf{G'}\vert \times \epsilon^{-1}(\mathbf{G},\mathbf{G'}).
\end{equation}
$\mathbf{k}$ is the Bloch vector restricted to the first Brillouin zone, $\mathbf{G}$ and $\mathbf{G'}$ are reciprocal lattice vectors, $n$ is the band index, $\omega$ is the frequency, $c$ is the speed of light; 
$\epsilon^{-1}(\mathbf{G},\mathbf{G'})$ is the Fourier transform of $\epsilon^{-1}(\mathbf{r})$. The dielectric function is invariant under translation by Bravais vectors $\mathbf{R}$, i.e. $\epsilon(\mathbf{r}+\mathbf{R})=\epsilon(\mathbf{r})$ and is derived from first principles simulations of the composing bulk materials. 
$c_n$ are the plane wave expansion coefficients of the magnetic field:
\begin{equation}
\mathbf{H}_{n,\mathbf{k}}(\mathbf{r})=\sum_{G}c_n(\mathbf{k}+\mathbf{G})\epsilon(\mathbf{k}+\mathbf{G})e^{i(\mathbf{k}+\mathbf{G})\cdot \mathbf{r}}.
\end{equation}
Similar formulation holds for the electric field.  

\section*{Supporting Data}
\begin{table*}[!t]
\caption{Effective Hubbard parameter ($U_{eff}$) calculated with ACBN0 code \cite{hubbard_dft, acbn0}; lattice parameter ($a_0$); and energy bandgap ($E_g$) of  III-V bulk semiconductors, simulated at the DFT+$U$ level of theory. The experimental reference values  are included for comparison.}
\label{tab:tabS1}
\begin{tabular}{|c|cc|cc|cc|}
\hline
\multirow{2}{*}{Material} & \multicolumn{2}{c|}{\multirow{2}{*}{$U_{eff}$ (eV)}} & \multicolumn{2}{c|}{$a_0$ (\AA)}                     & \multicolumn{2}{c|}{$E_g$ (eV)}                                     \\ \cline{4-7} 
                          & \multicolumn{2}{c|}{}                                & \multicolumn{1}{l|}{This work} & \multicolumn{1}{l|}{Reference$^a$} & \multicolumn{1}{l|}{This work} & \multicolumn{1}{l|}{Reference$^a$} \\ \hline
AlP                       & \multicolumn{1}{c|}{Al(3p)=16.88}    & P (3p)=2.36   & \multicolumn{1}{c|}{5.485}     & 5.451                              & \multicolumn{1}{c|}{1.88}      & 2.50                               \\ 
AlAs                      & \multicolumn{1}{c|}{Al (3p)=16.87}   & As (4p)=2.08  & \multicolumn{1}{c|}{5.692}     & 5.661                              & \multicolumn{1}{c|}{1.67}      & 2.12                               \\ 
AlSb                      & \multicolumn{1}{c|}{Al (3p)=16.87}   & Sb (5p)=1.51  & \multicolumn{1}{c|}{6.169}     & 6.136                              & \multicolumn{1}{c|}{1.28}      & 1.60                               \\ 
GaP                       & \multicolumn{1}{c|}{Ga (3d)=19.56}   & P (3p)=1.96   & \multicolumn{1}{c|}{5.483}     & 5.451                              & \multicolumn{1}{c|}{1.80}      & 2.27                               \\ 
GaAs                      & \multicolumn{1}{c|}{Ga (3d)=19.61}   & As (4p)=1.75  & \multicolumn{1}{c|}{5.688}     & 5.653                              & \multicolumn{1}{c|}{0.74}      & 1.42                               \\ 
GaSb                      & \multicolumn{1}{c|}{Ga (3d)=19.47}   & Sb (5p)=1.30  & \multicolumn{1}{c|}{6.147}     & 6.096                              & \multicolumn{1}{c|}{0.24}      & 0.75                               \\ 
InP                       & \multicolumn{1}{c|}{In (4d)=14.23}   & P (3p)=2.18   & \multicolumn{1}{c|}{5.912}     & 5.869                              & \multicolumn{1}{c|}{1.08}      & 1.42                               \\ 
InAs                      & \multicolumn{1}{c|}{In (4d)=14.30}   & As (4p)=1.94  & \multicolumn{1}{c|}{6.151}     & 6.058                              & \multicolumn{1}{c|}{0.25}      & 0.42                               \\ 
InSb                      & \multicolumn{1}{c|}{In (4d)=14.37}   & Sb (5p)=1.45  & \multicolumn{1}{c|}{6.578}     & 6.479                              & \multicolumn{1}{c|}{0.18}      & 0.24                               \\ \hline
\end{tabular}\\
$^a$ Data from Ref. [\citenum{singh1993}].
\end{table*}

\begin{figure*}[]
	\centering
	\includegraphics[width=0.6\linewidth]{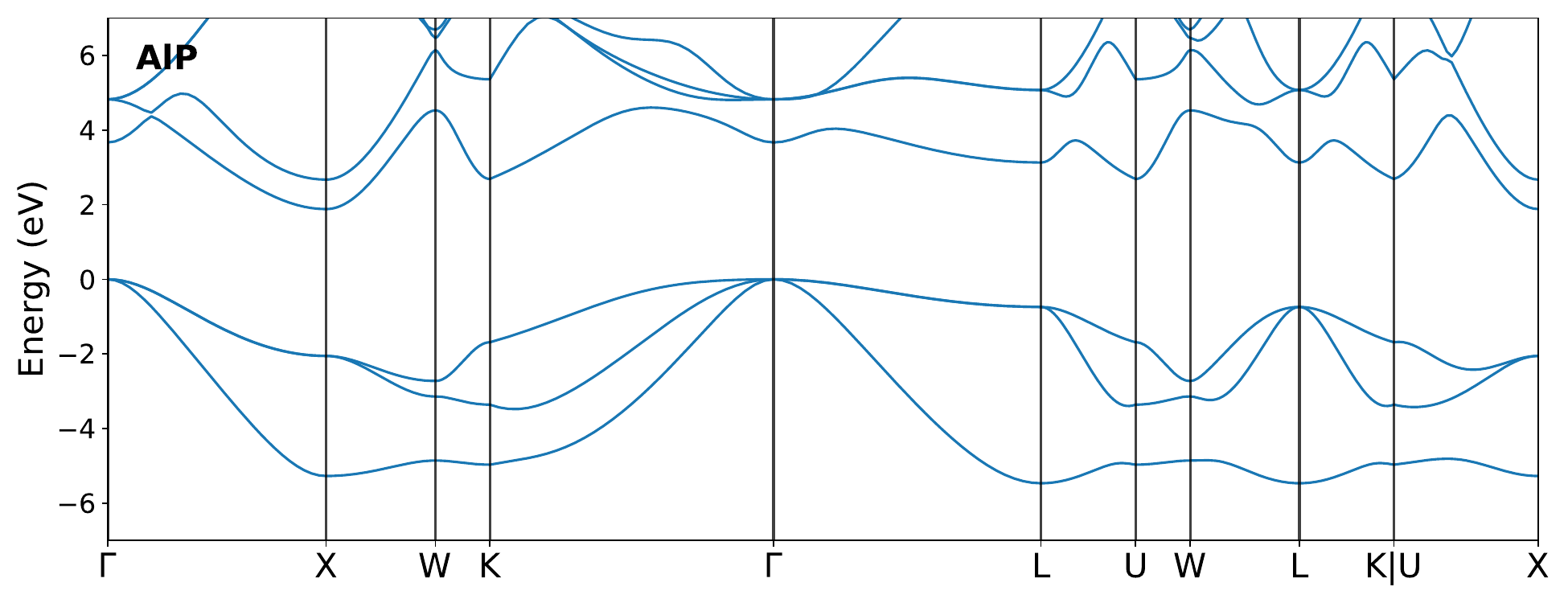} \\ 
	\includegraphics[width=0.6\linewidth]{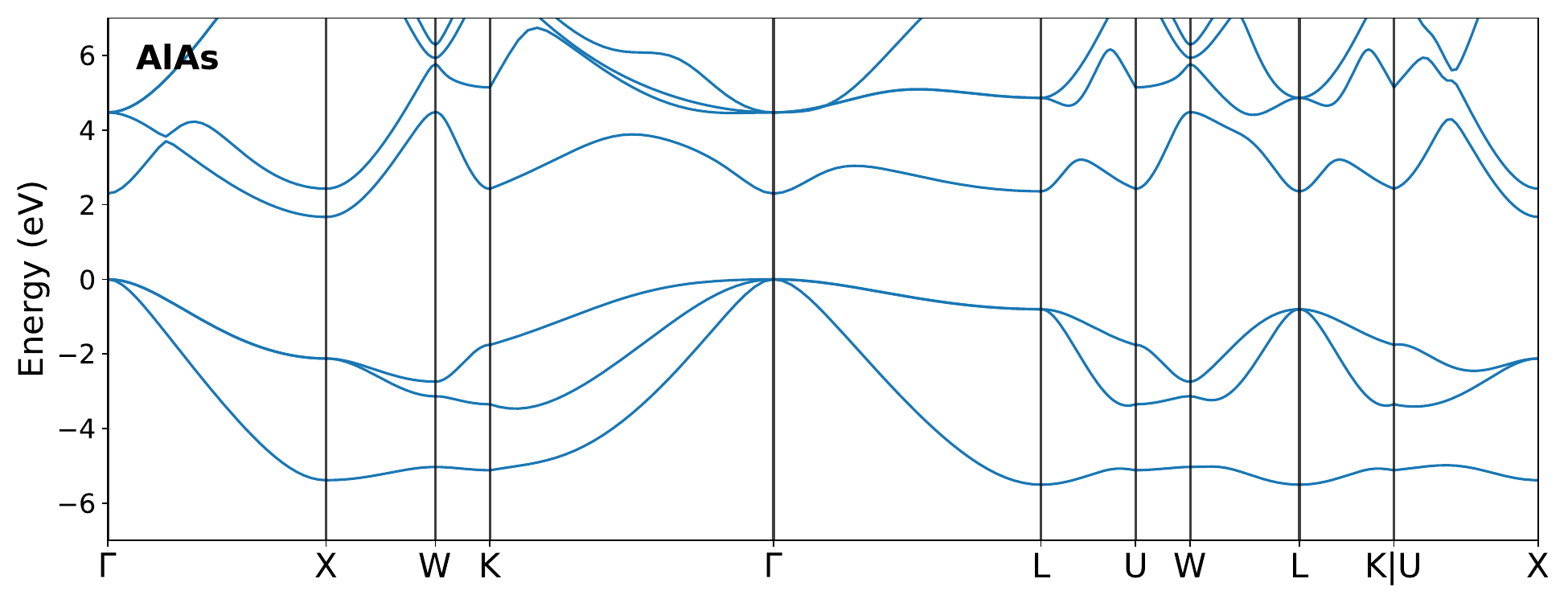} \\
	\includegraphics[width=0.6\linewidth]{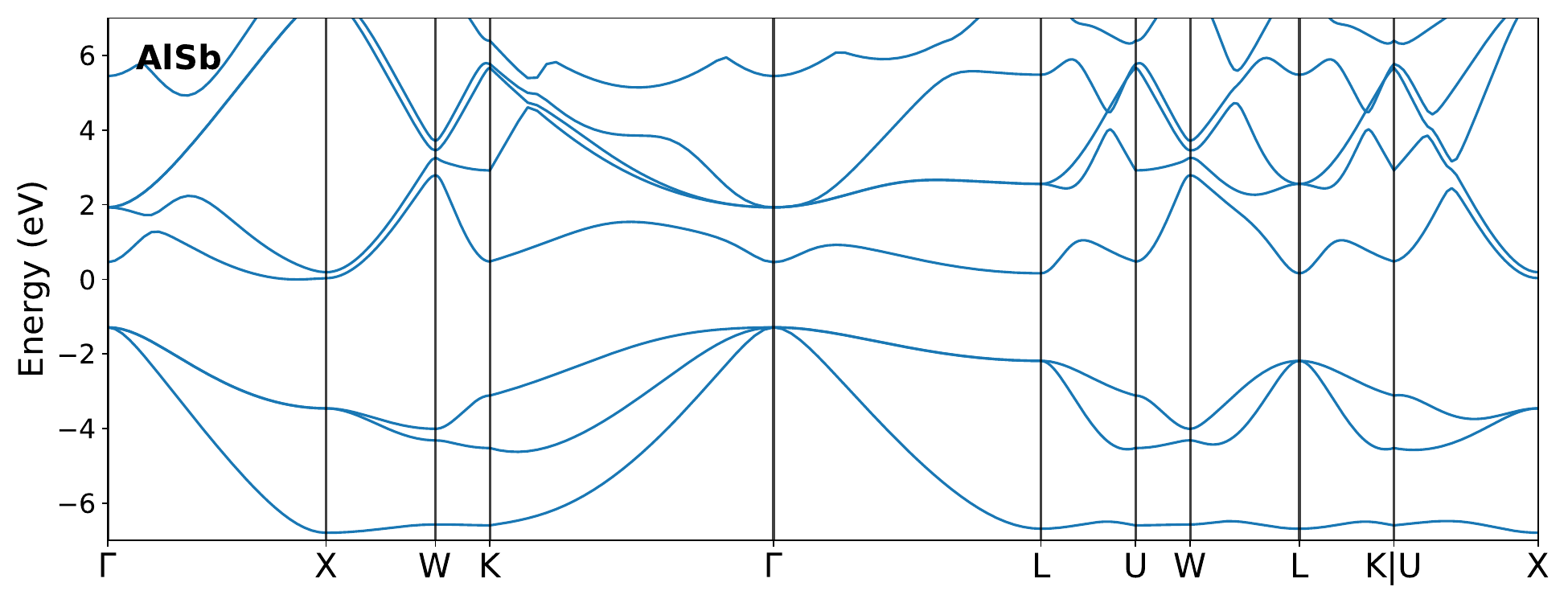} 			
	\caption{Energy band structures of AlP (upper panel), AlAs (middle panel), and AlSb (lower panel), calculated at the DFT+$U$ level of theory. The zero is set to the Fermi level of each system.}
	\label{bande1}
\end{figure*}

\begin{figure*}[]
	\centering
	\includegraphics[width=0.6\linewidth]{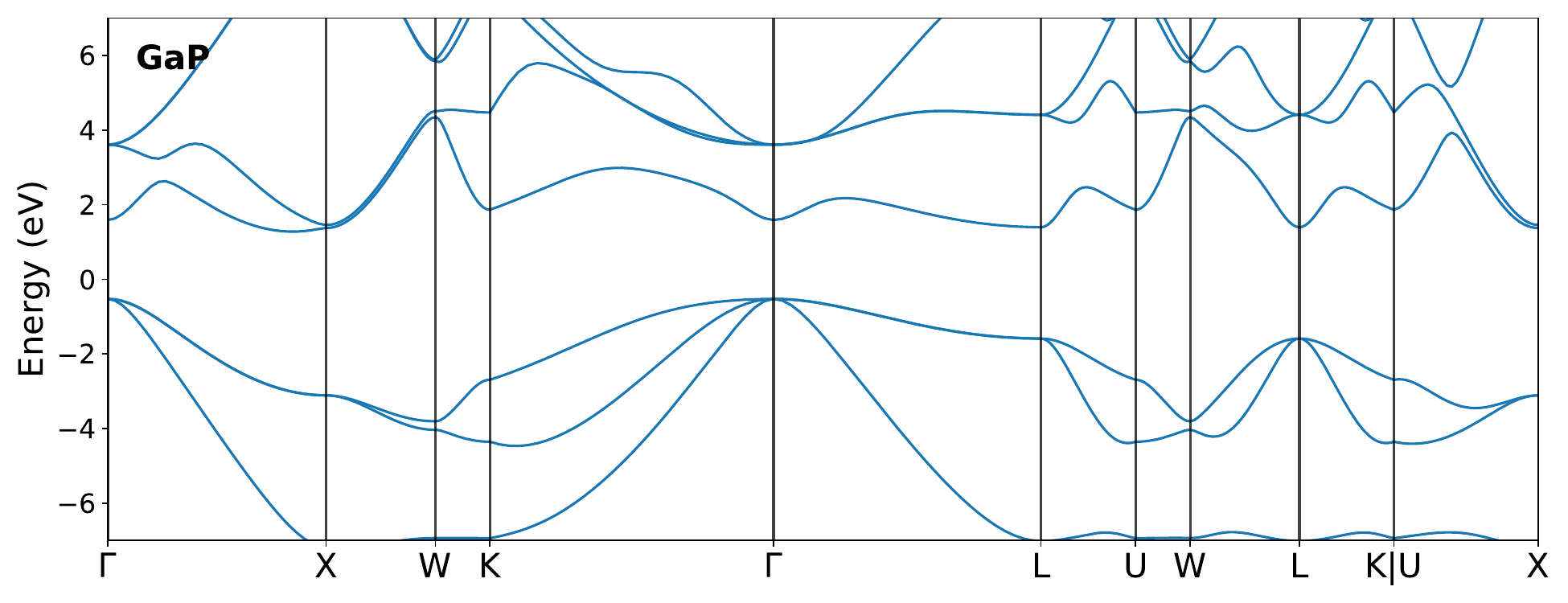} \\	
	\includegraphics[width=0.6\linewidth]{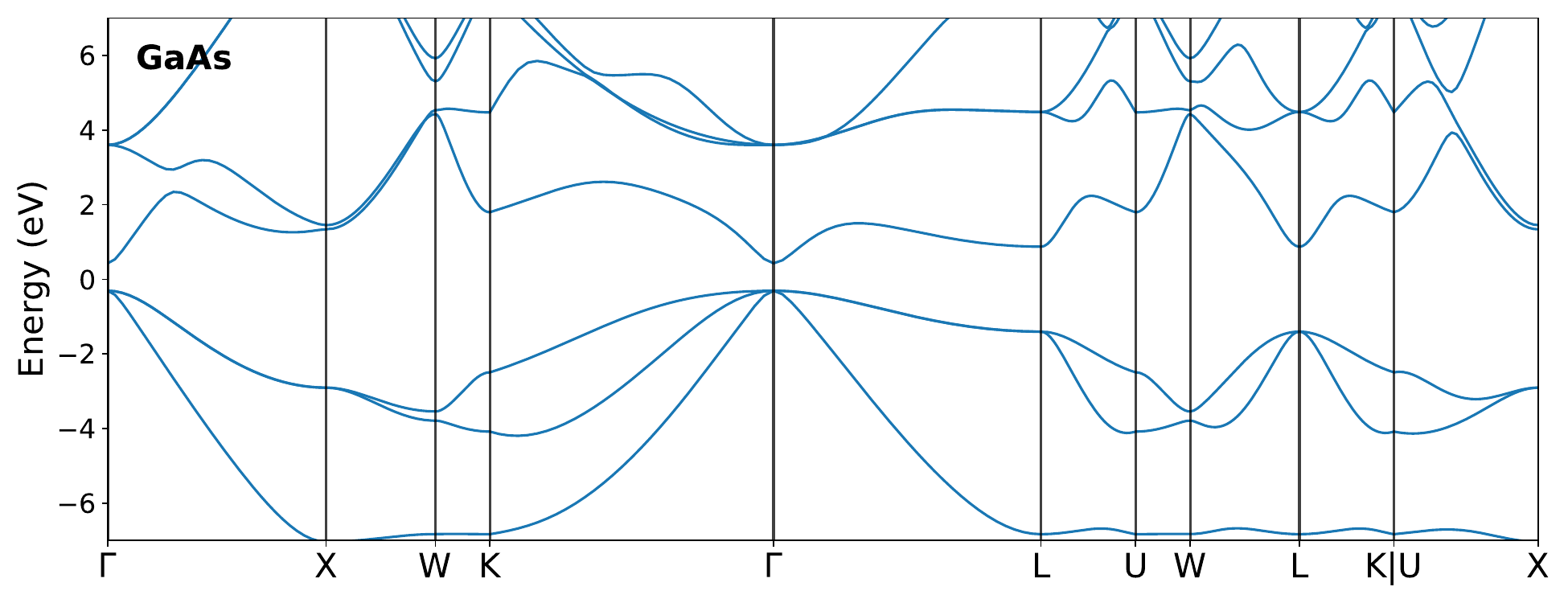}  \\
	\includegraphics[width=0.60\linewidth]{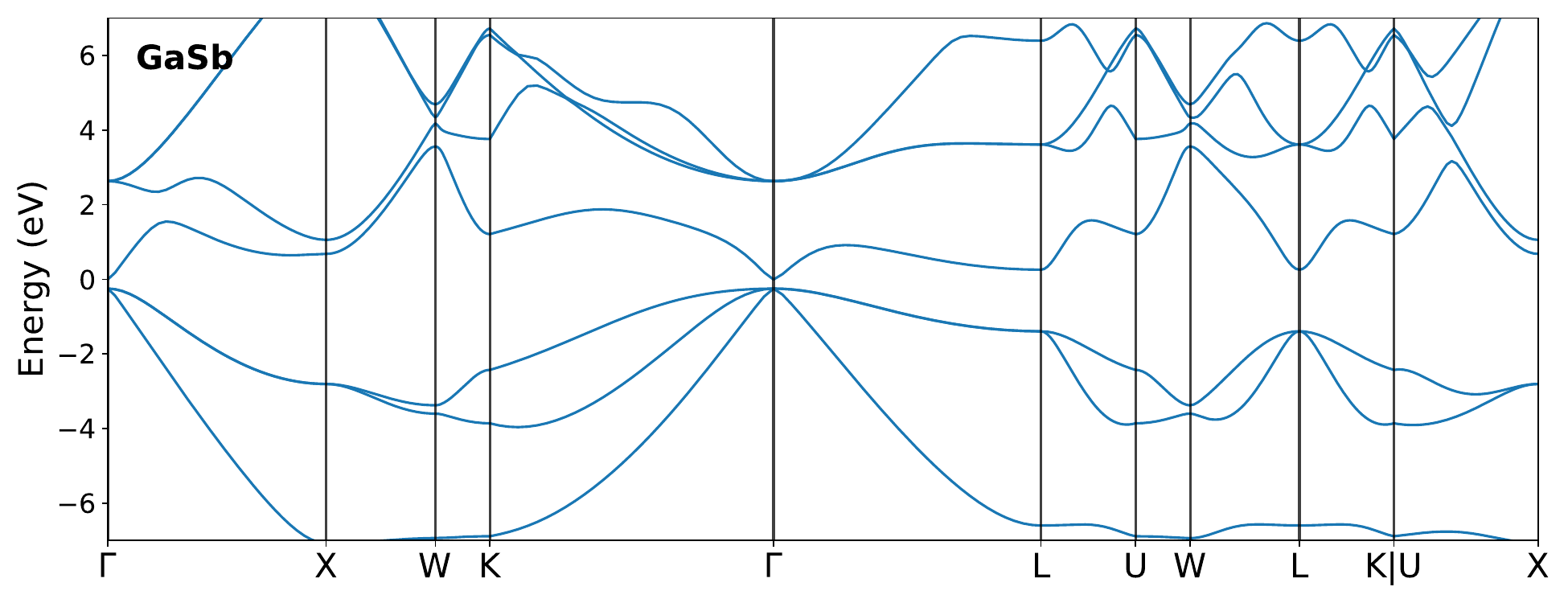} 
	\caption{Energy band structures of GaP (upper panel), GaAs (middle panel), and GaSb (lower panel), calculated at the DFT+$U$ level of theory. The zero is set to the Fermi level of each system.}
	\label{bande2}
\end{figure*}

\begin{figure*}[]
	\centering
	\includegraphics[width=0.6\linewidth]{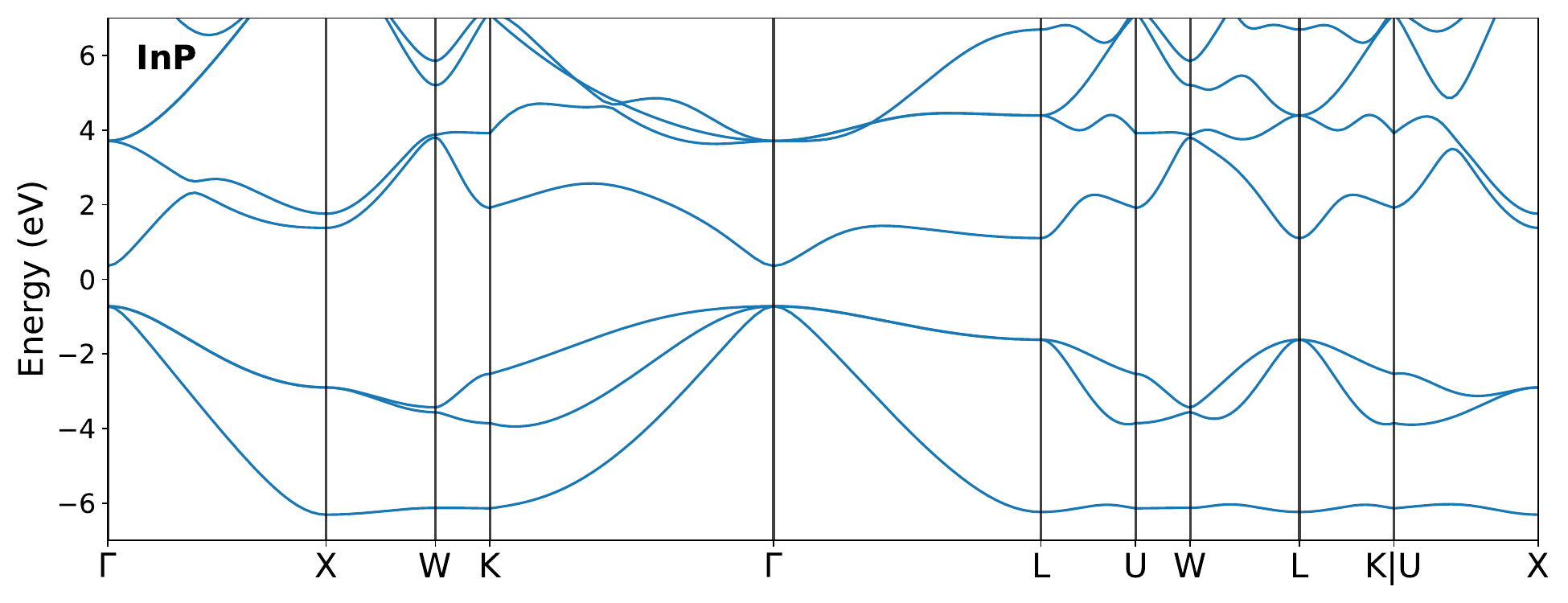} \\
	\includegraphics[width=0.6\linewidth]{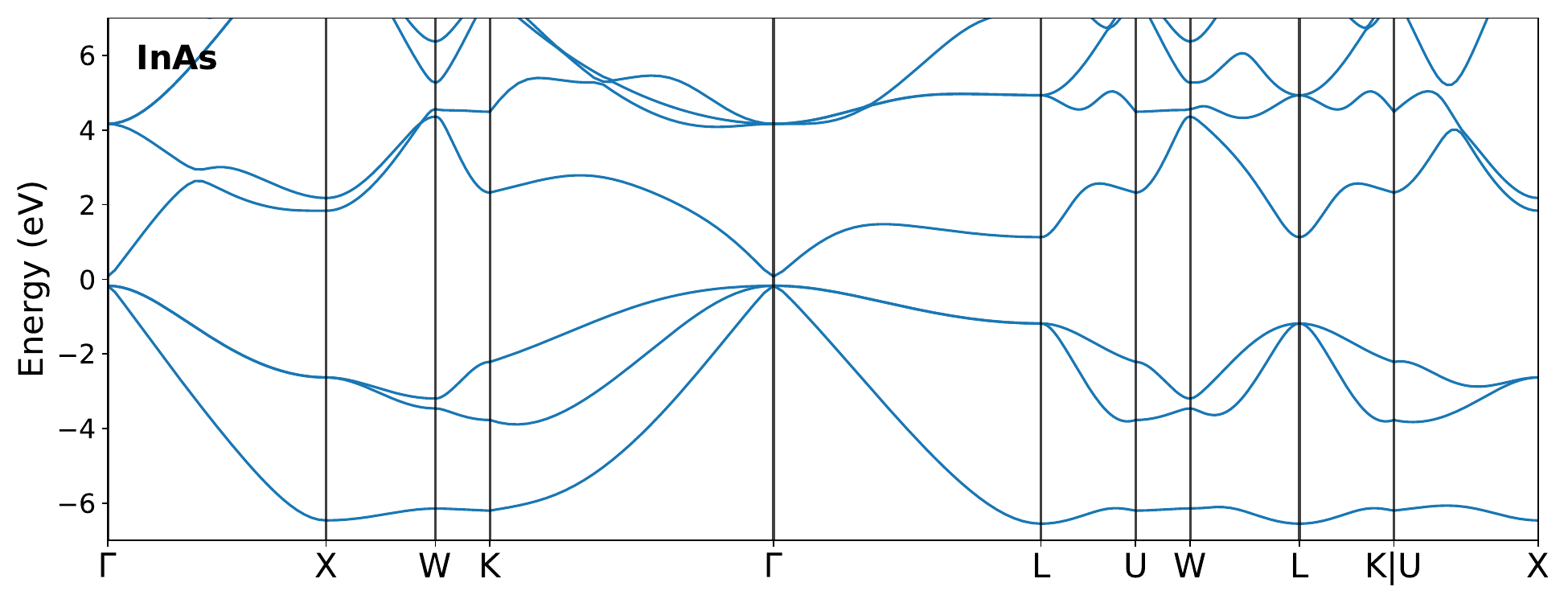} \\		
	\includegraphics[width=0.6\linewidth]{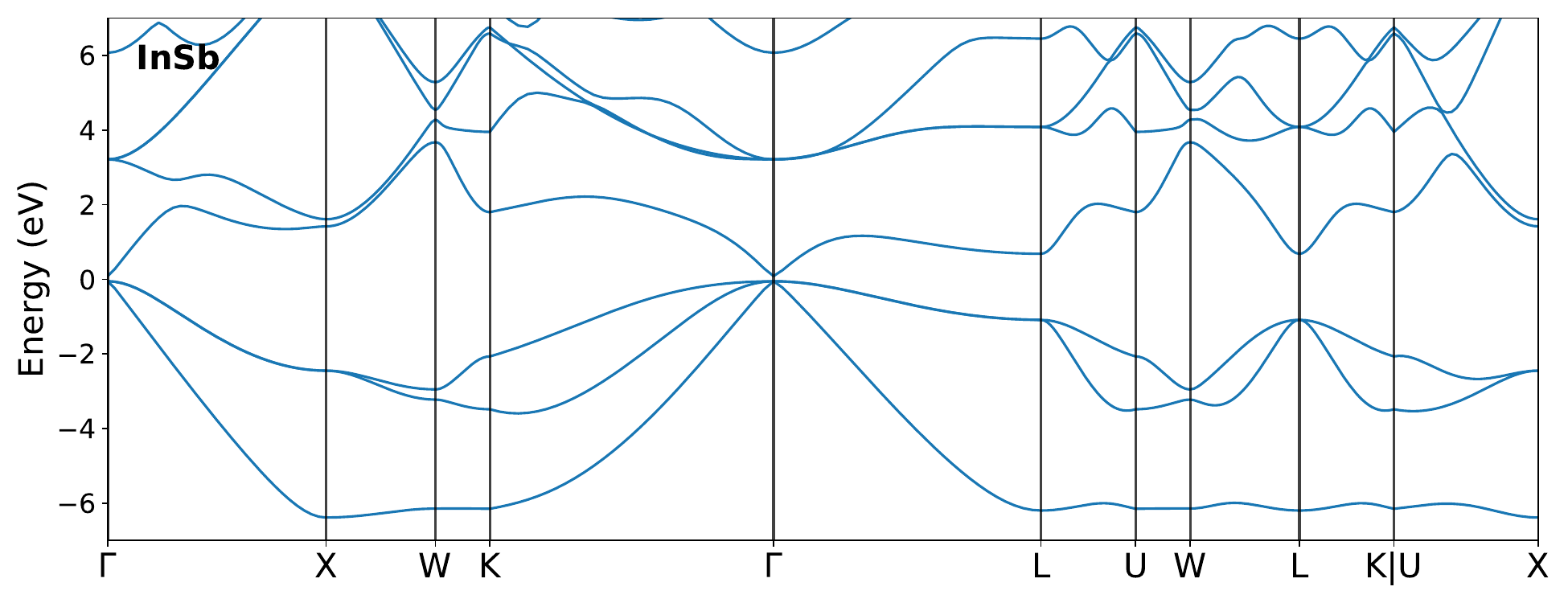}  	
	\caption{Energy band structures of InP (upper panel), InAs (middle panel), and InSb (lower panel), calculated at the DFT+$U$ level of theory. The zero is set to the Fermi level of each system. }
	\label{bande3}
\end{figure*}

\begin{figure*}[]
\centering
\includegraphics[width=0.5\linewidth]{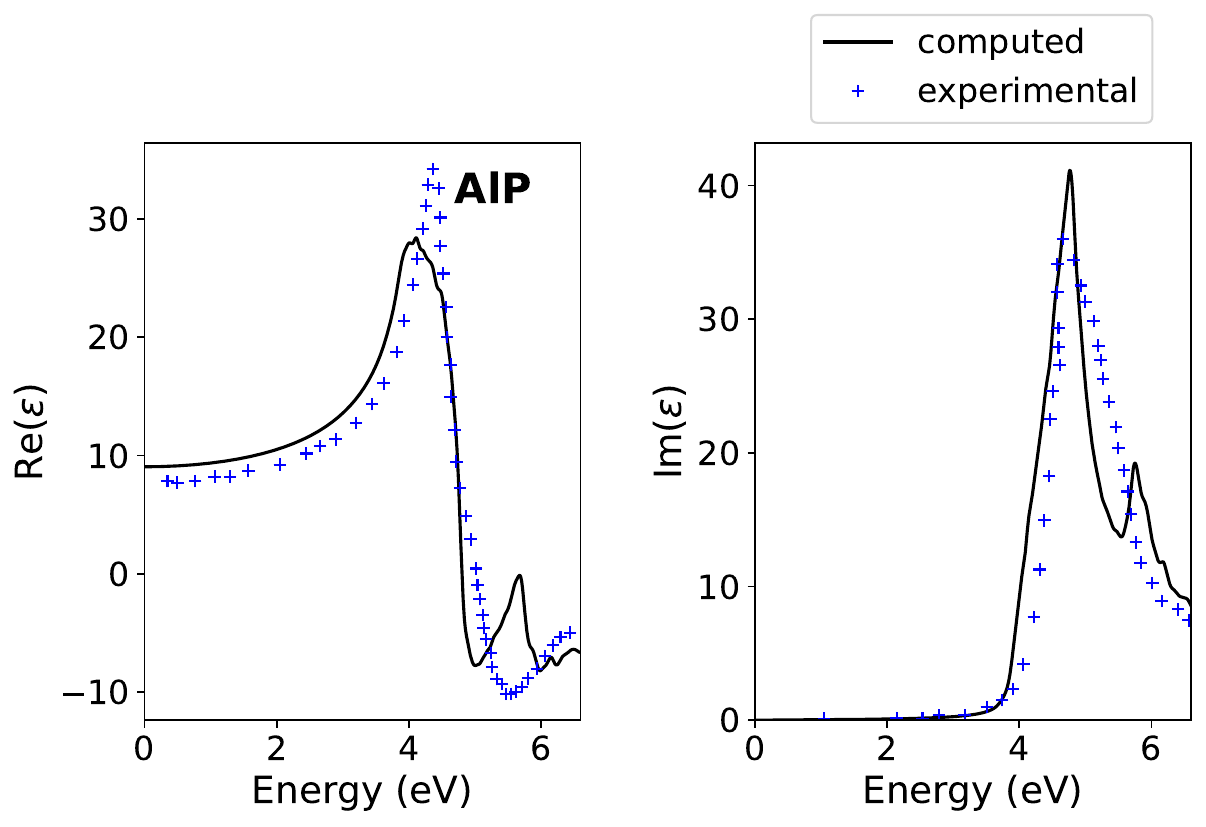}\\
\includegraphics[width=0.5\linewidth]{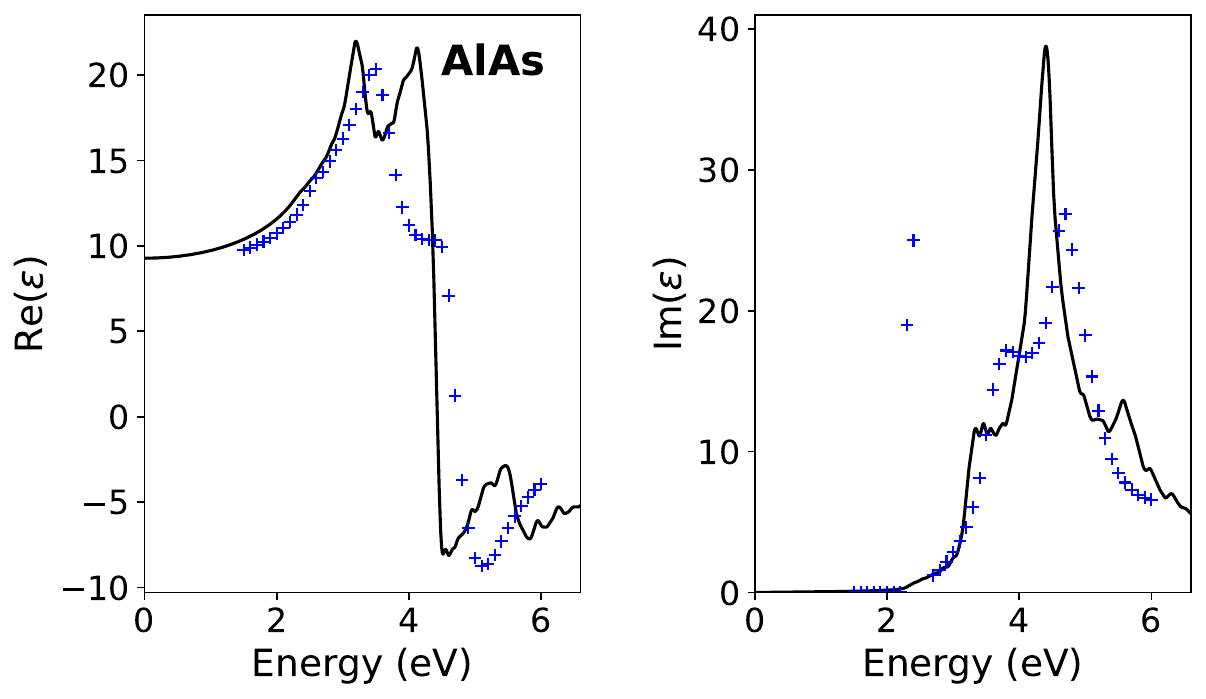}\\
\includegraphics[width=0.5\linewidth]{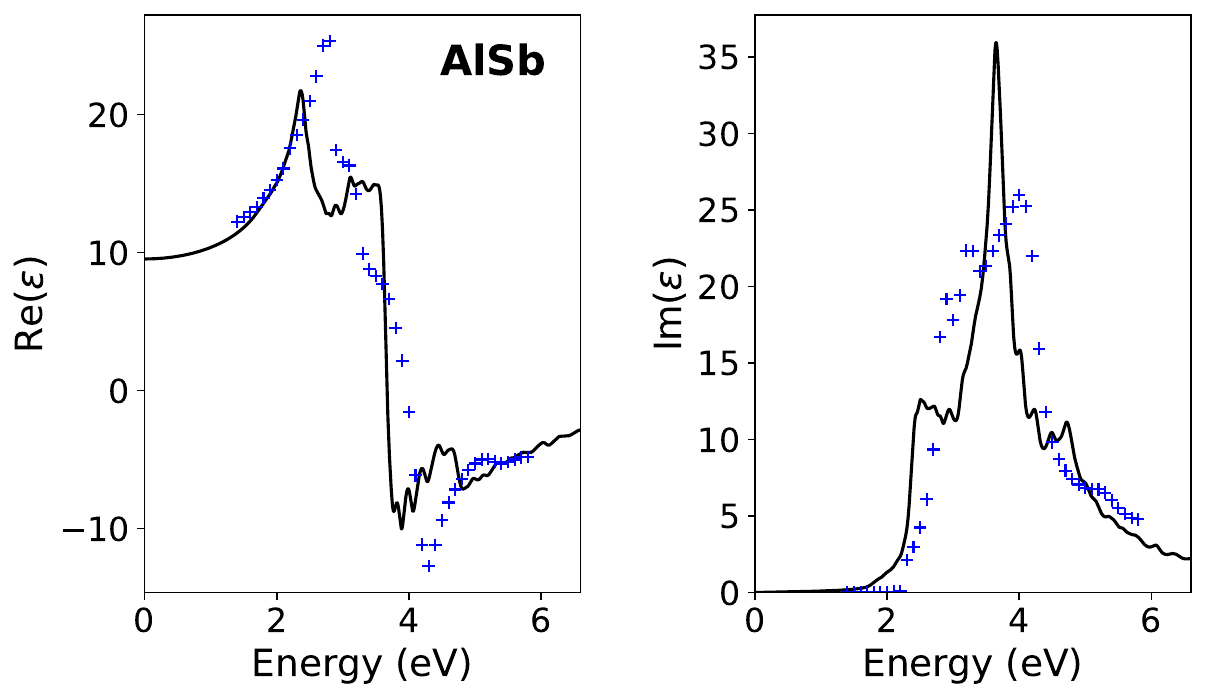}
\caption{Comparison between the experimental (blue cross symbols) and simulated (black lines) real (left) and imaginary (right) parts of the dielectric functions of (a) AlP, (b) AlAs, and (c) AlSb bulk semiconductors. 
Reference experimental data are adapted from Refs. [\citenum{epsm_AlP,epsm_AlAs,epsm_AlSb}], respectively.}
\label{eps1}
\end{figure*}

\begin{figure*}[]
\centering
\includegraphics[width=0.5\linewidth]{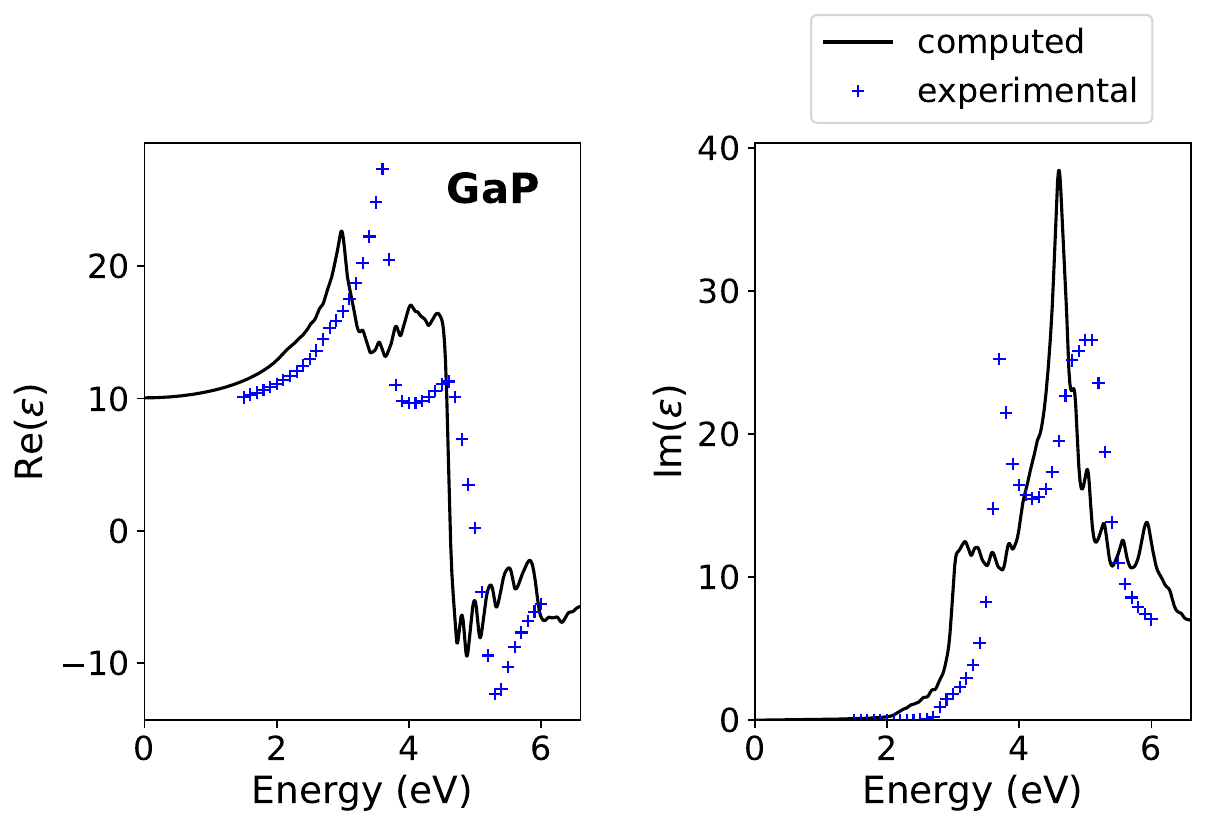} \\
\includegraphics[width=0.5\linewidth]{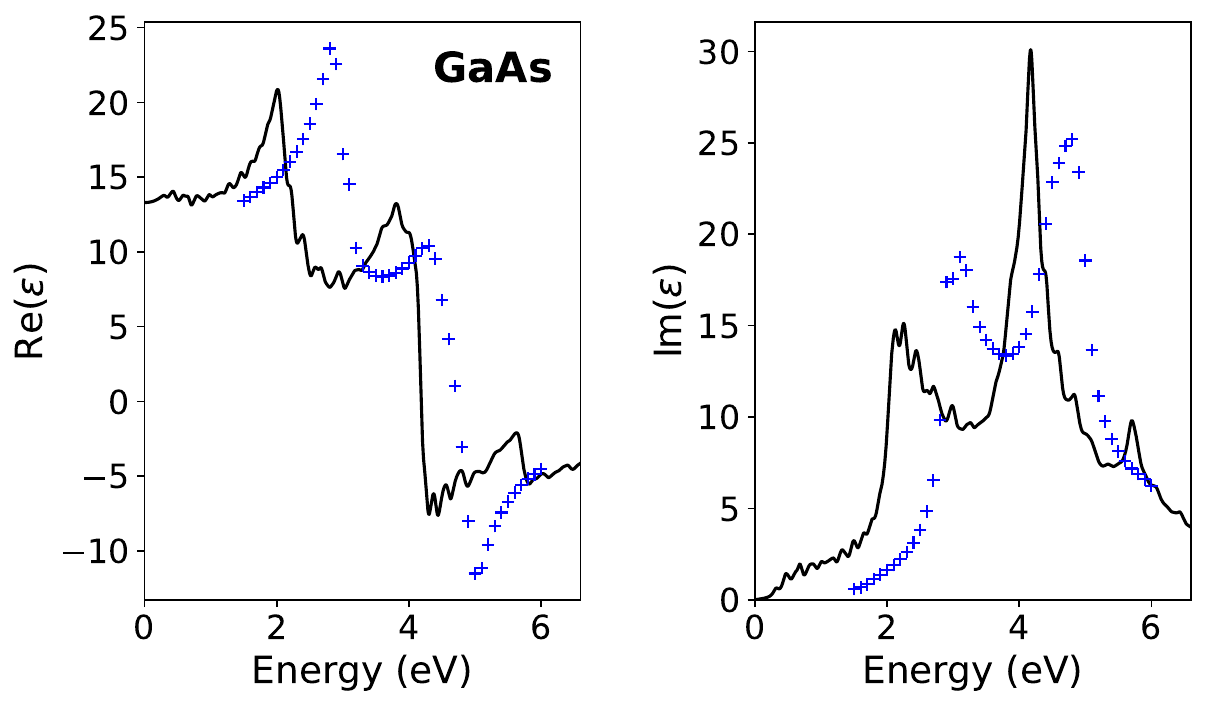}\\
\includegraphics[width=0.5\linewidth]{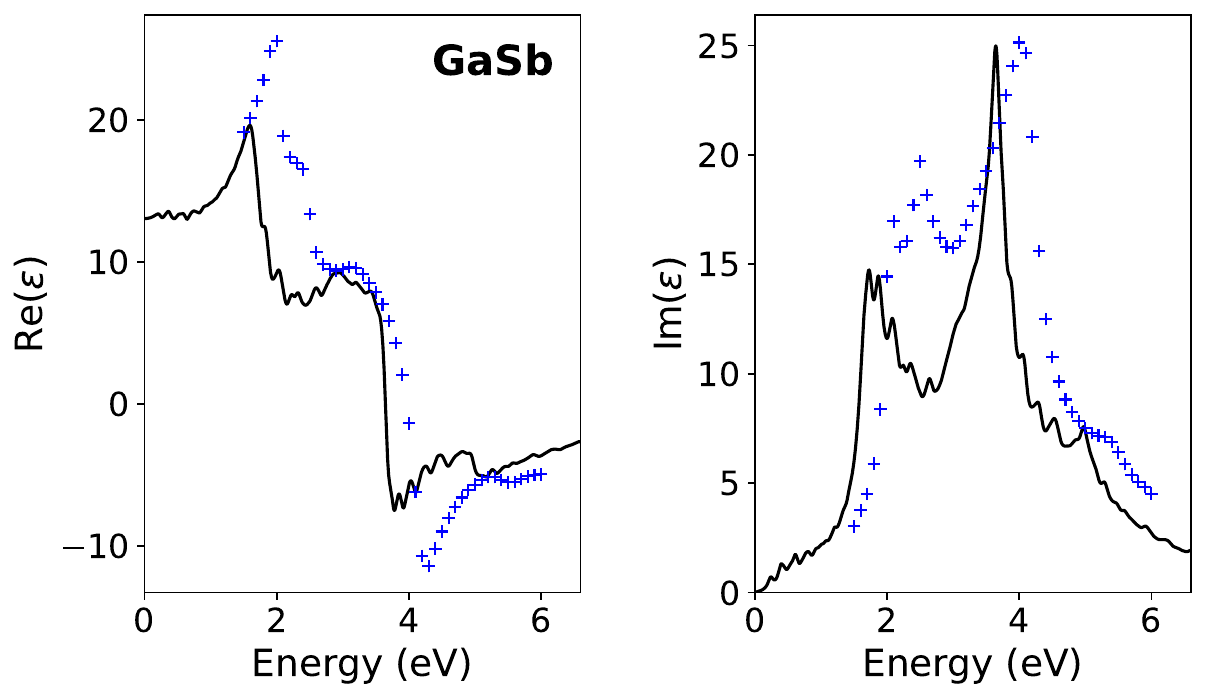}
\caption{Comparison between the experimental (blue cross symbols) and simulated (black lines) real (left) and imaginary (right) parts of the dielectric functions of (a) GaP, (b) GaAs, and (c) GaSb bulk semiconductors. Reference experimental data are adapted from Ref. [\citenum{dielectric1983}].} 
\label{eps2}
\end{figure*}

\begin{figure*}[]
\centering
\includegraphics[width=0.5\linewidth]{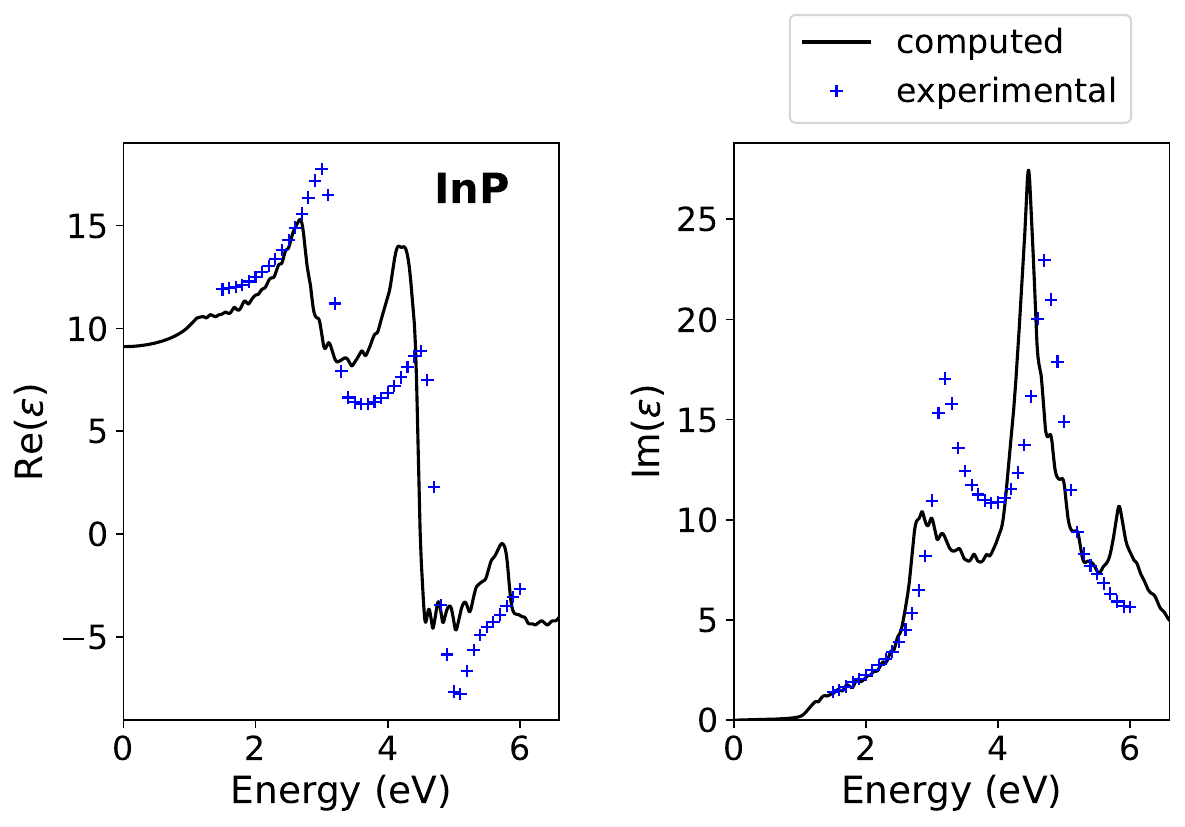} \\
\includegraphics[width=0.5\linewidth]{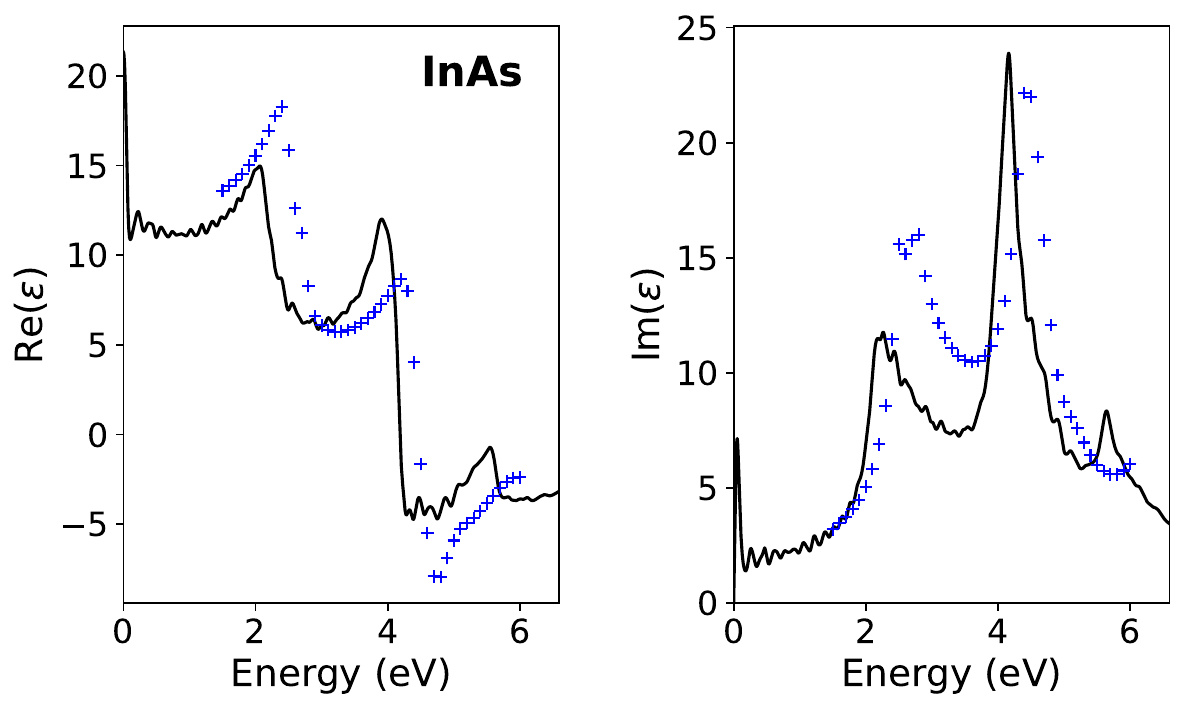} \\
\includegraphics[width=0.5\linewidth]{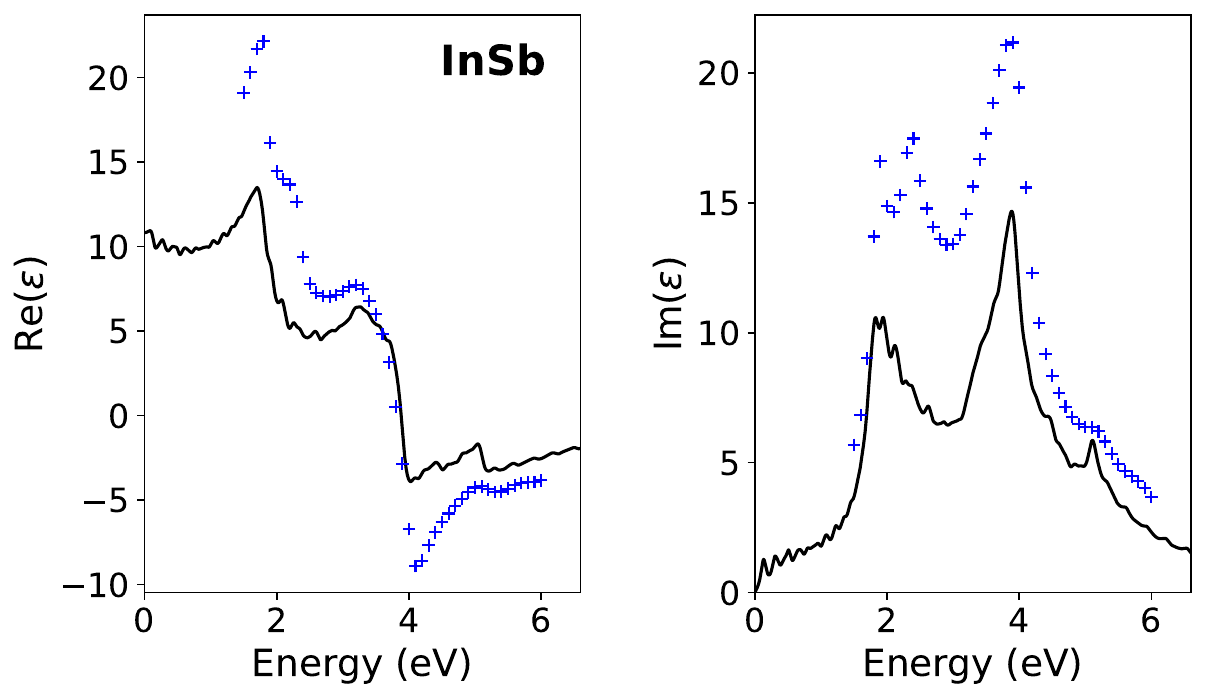}
\caption{Comparison between the experimental (blue cross symbols) and simulated (black lines) real (left) and imaginary (right) parts of the dielectric functions of (a) InP, (b) InAs, and (c) InSb bulk semiconductors. Reference experimental data are adapted from Ref. [\citenum{dielectric1983}].} 
\label{eps3}
\end{figure*}

\begin{figure*}[]
\centering
\includegraphics[width=0.45\linewidth]{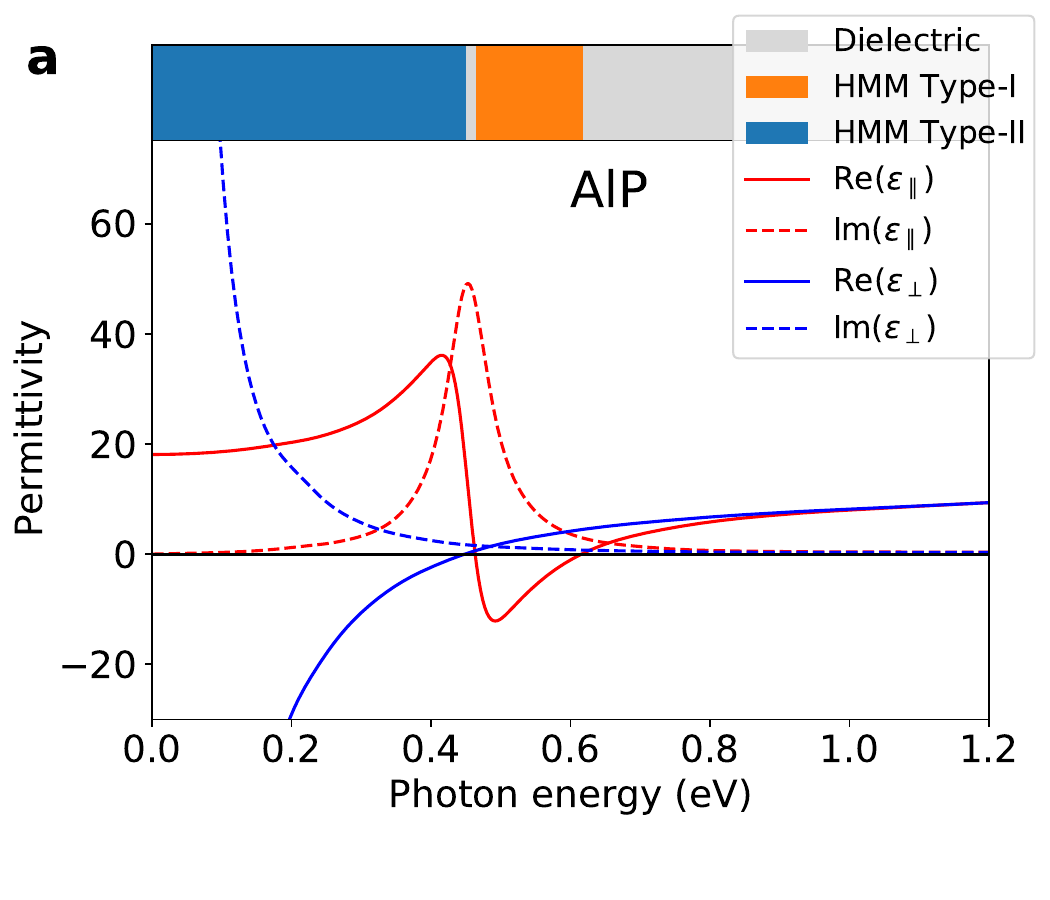} \\
\includegraphics[width=0.45\linewidth]{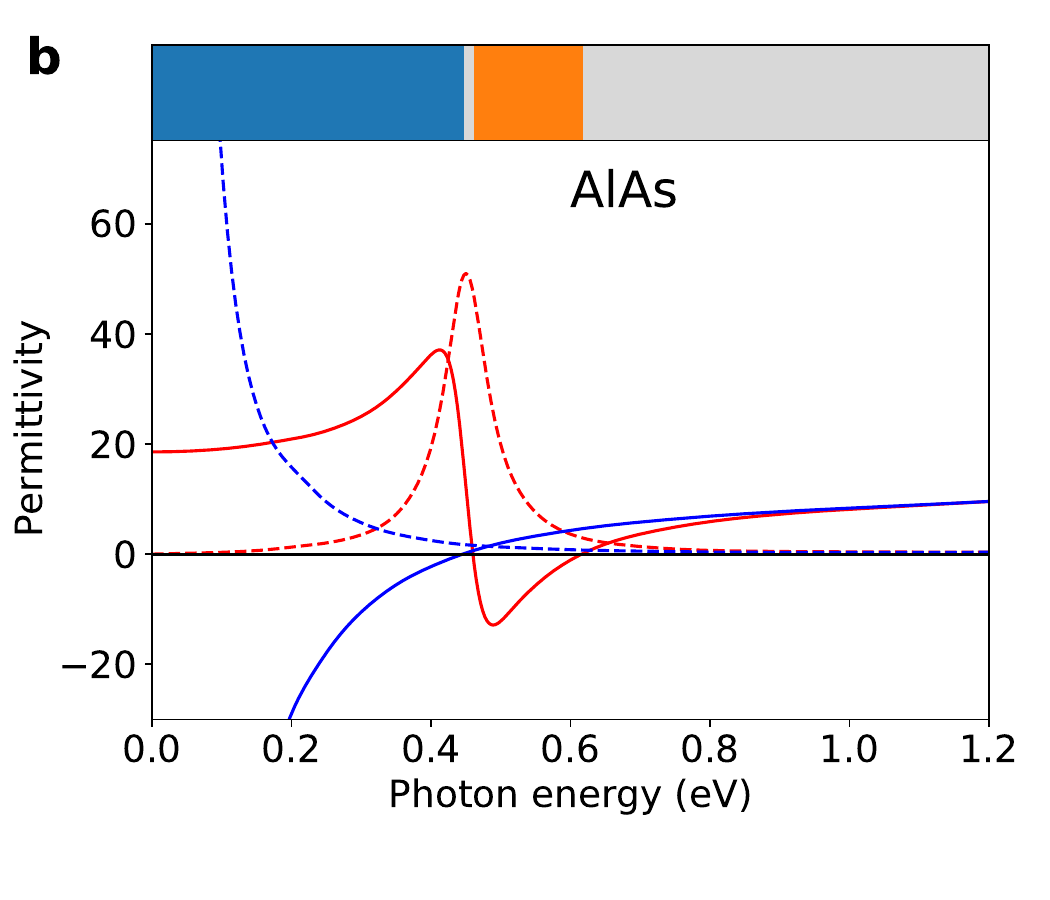} \\
\includegraphics[width=0.45\linewidth]{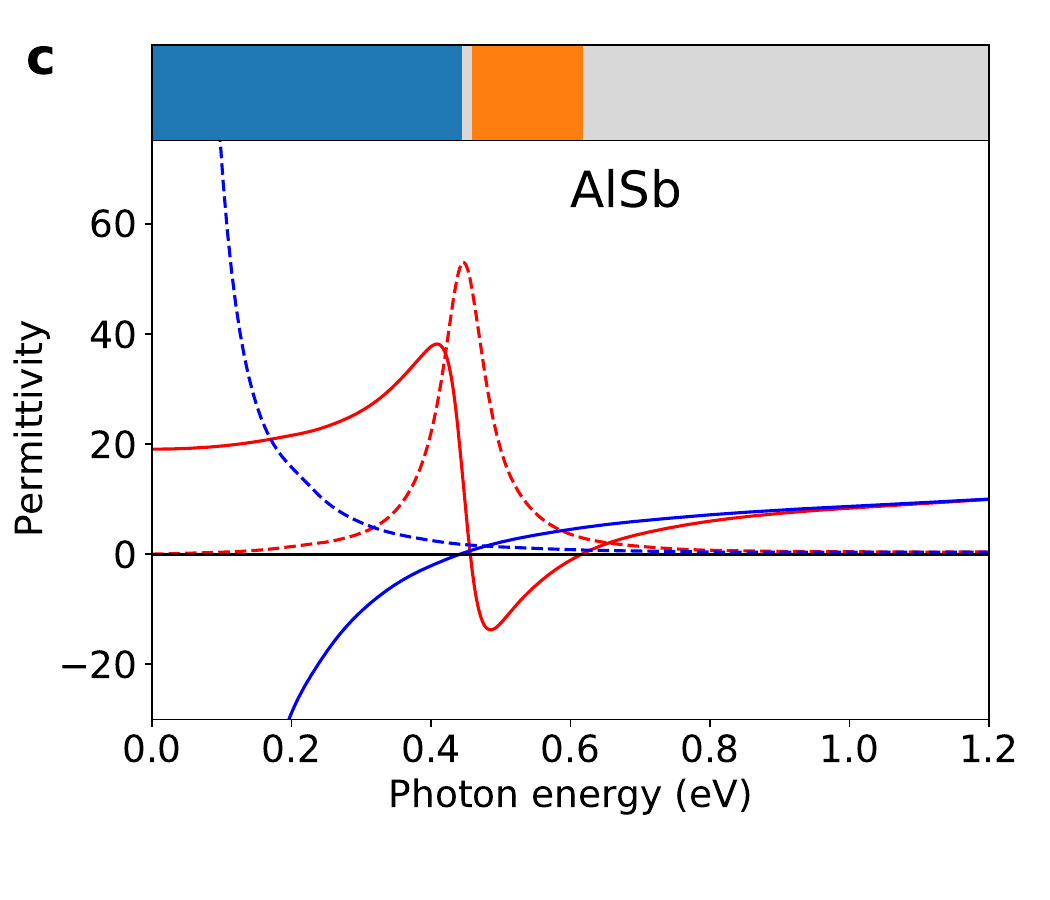}
\caption{Real (solid lines) and imaginary (dashed lines) parts of the parallel (red lines) and perpendicular (blue lines) dielectric functions of metamaterials composed of AlX (X = P (a), As (b), Sb (c)) and Si:InAs ($c_{\bf 2}=$3.12\%) with equal filling factors ($f_1=0.5$).
The bars above the graphs indicate the energy regions with type-I (orange), and type-II (blue) hyperbolic, and dielectric (gray) behaviors. Legend is provided in panel (a).}
\label{mat1}
\end{figure*}

\begin{figure*}[]
\centering
\includegraphics[width=0.45\linewidth]{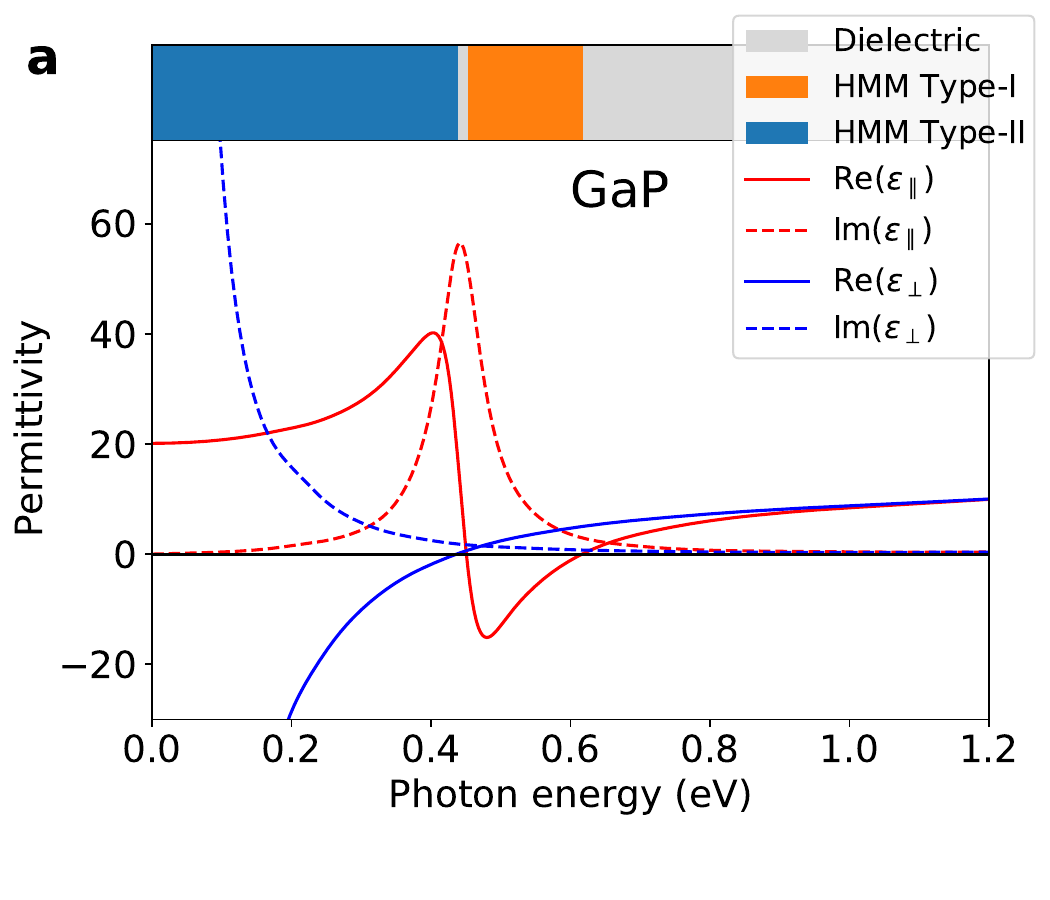} \\
\includegraphics[width=0.45\linewidth]{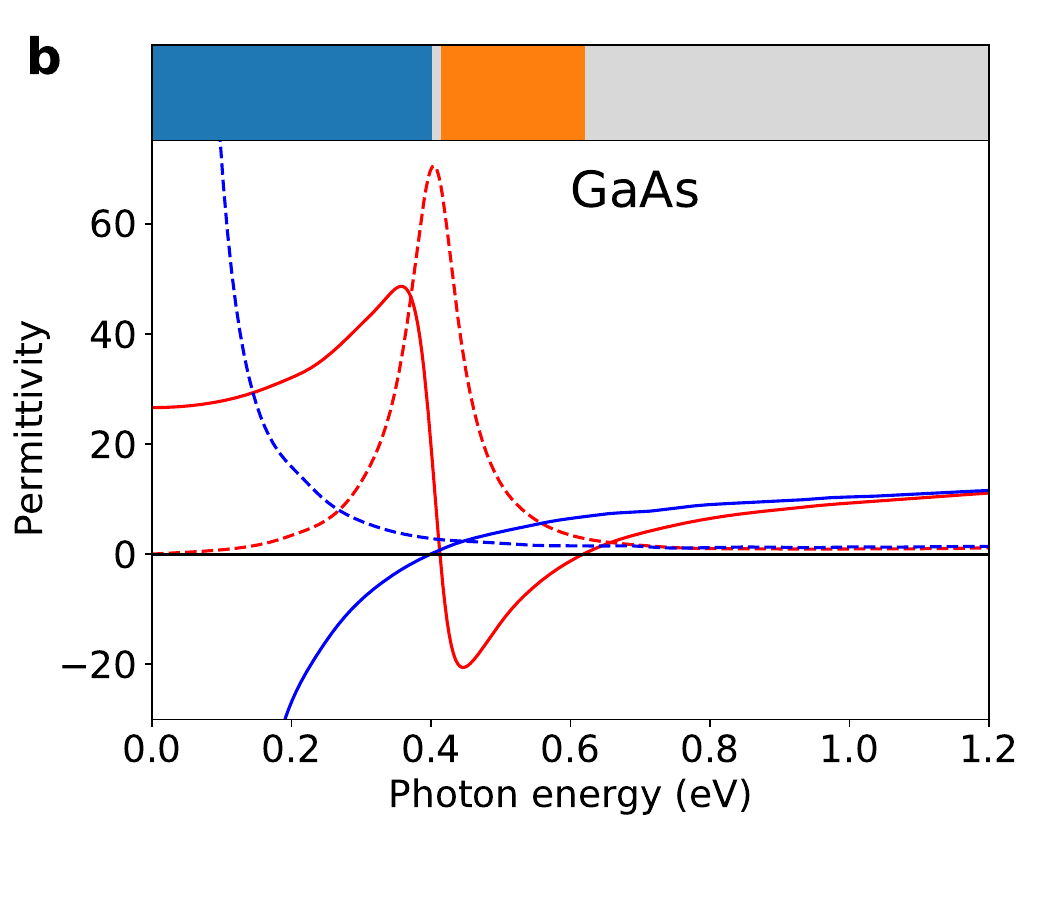} \\
\includegraphics[width=0.45\linewidth]{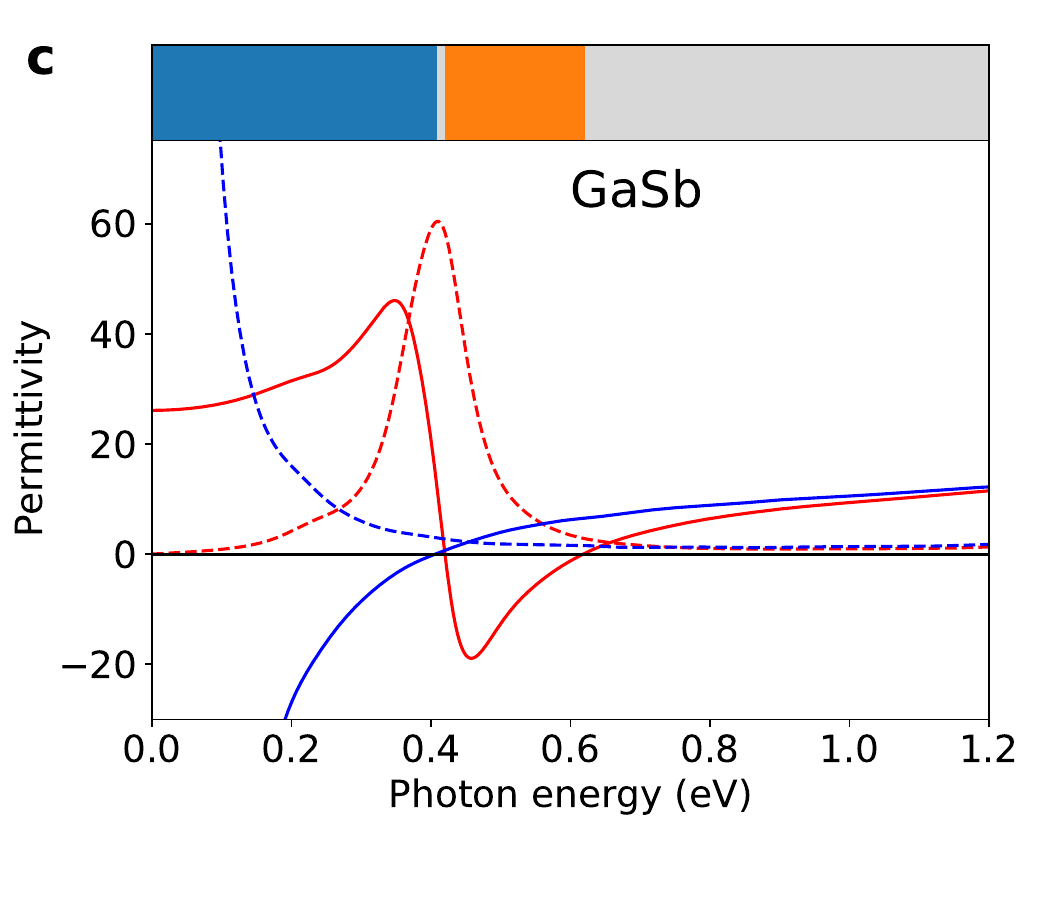}
\caption{Real (solid lines) and imaginary (dashed lines) parts of the parallel (red lines) and perpendicular (blue lines) dielectric functions of metamaterials composed of GaX (X = P (a), As (b), Sb (c)) and Si:InAs ($c_{\bf 2}=$3.12\%)  with equal filling factors ($f_1=0.5$).
The bars above the graphs indicate the energy regions with type-I (orange), and type-II (blue) hyperbolic, and dielectric (gray) behaviors. Legend is provided in panel (a).}
\label{mat2}
\end{figure*}

\begin{figure*}[]
\centering
\includegraphics[width=0.45\linewidth]{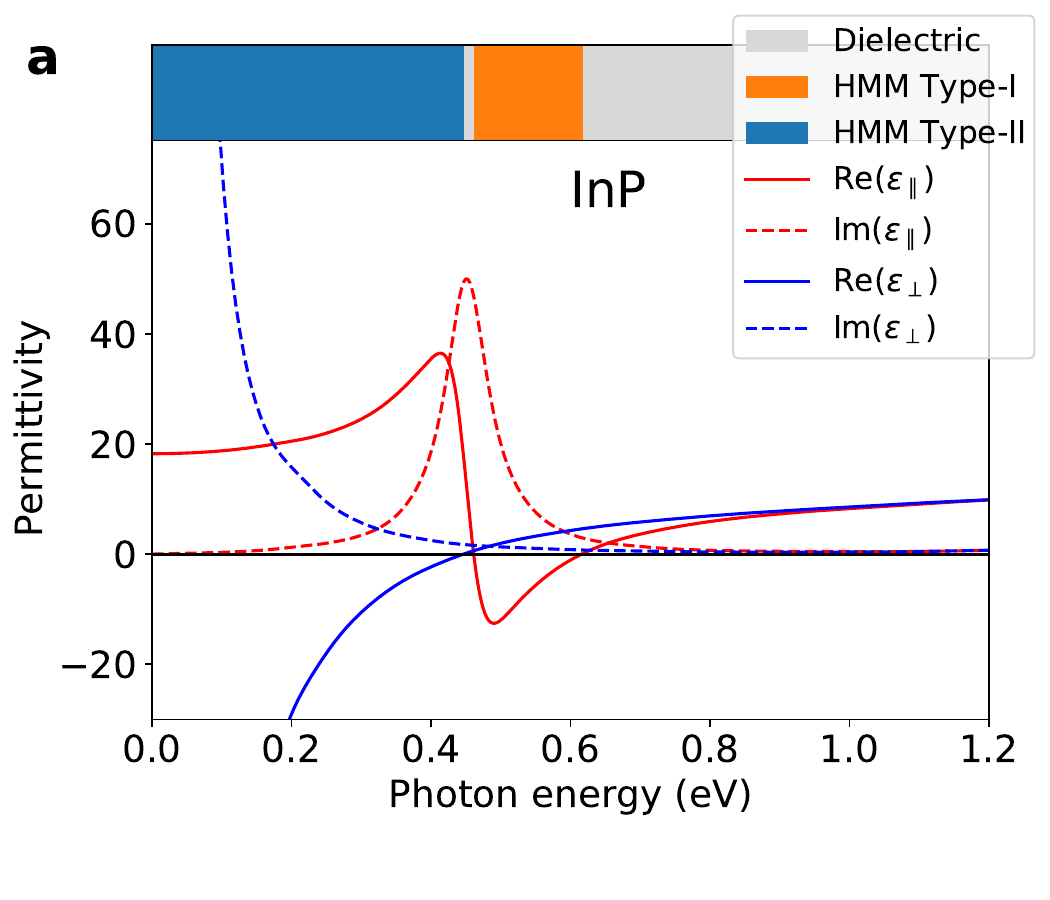} \\
\includegraphics[width=0.45\linewidth]{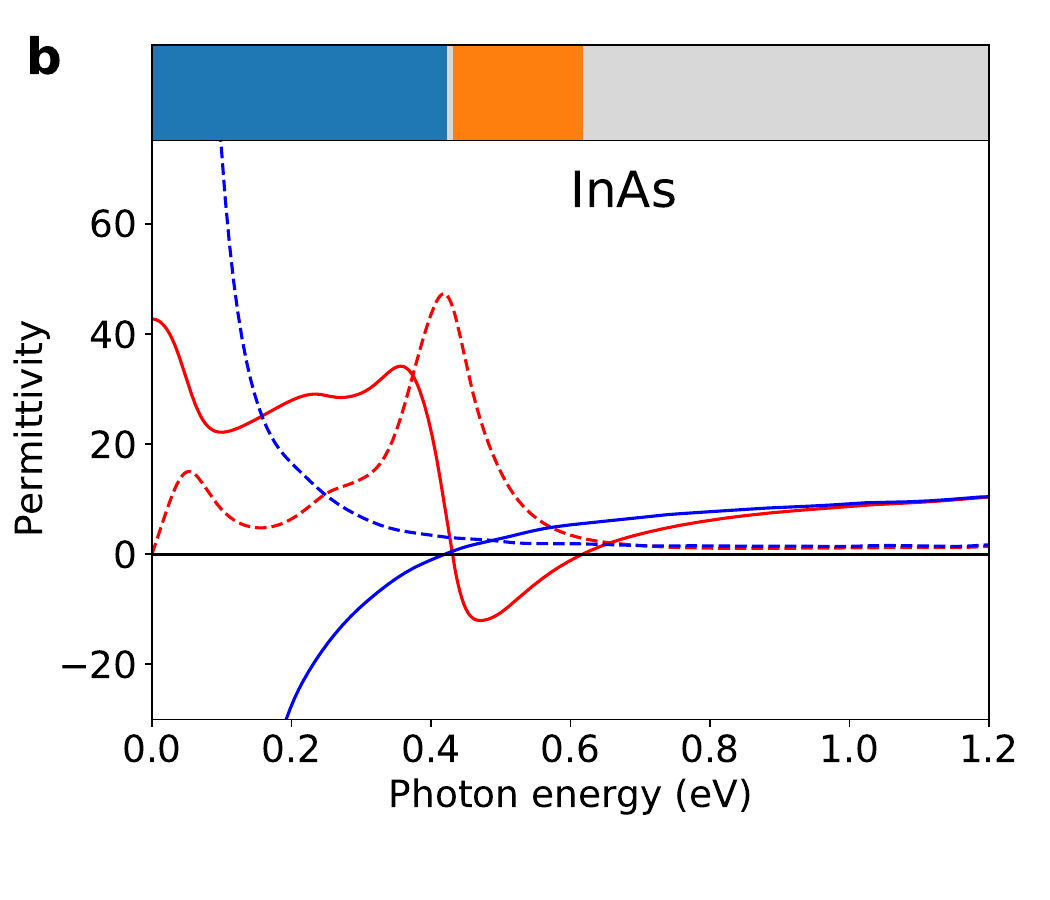} \\
\includegraphics[width=0.45\linewidth]{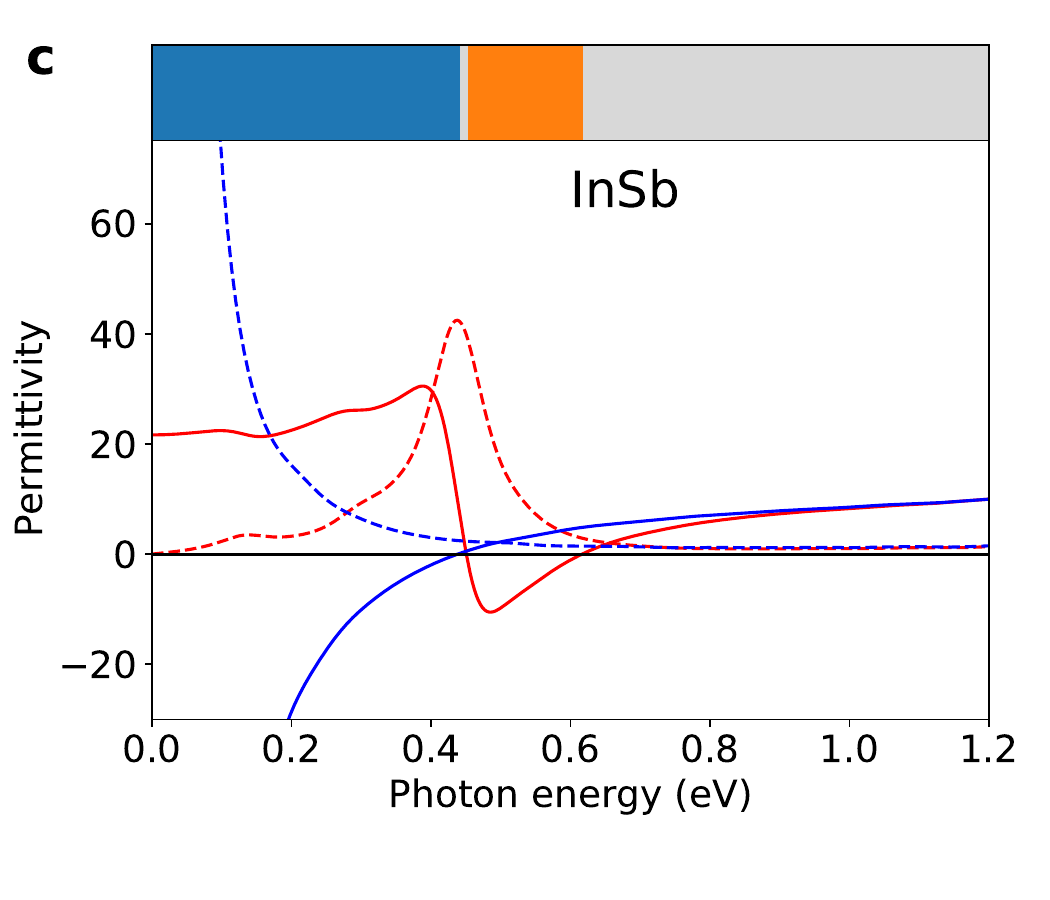}\\
\caption{Real (solid lines) and imaginary (dashed lines) parts of the parallel (red lines) and perpendicular (blue lines) dielectric functions of metamaterials composed of InX (X = P (a), As (b), Sb (c)) and Si:InAs ($c_{\bf 2}=$3.12\%)  with equal filling factors ($f_1=0.5$).
The bars above the graphs indicate the energy regions with type-I (orange), and type-II (blue) hyperbolic, and dielectric (gray) behaviors. Legend is provided in panel (a).}
\label{mat3}
\end{figure*}

\begin{figure*}[h!]
\centering
\includegraphics[width=0.6\linewidth]{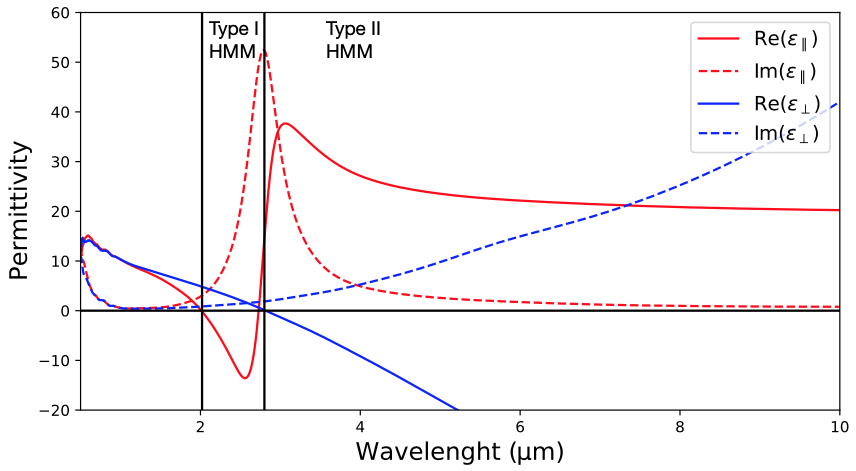}
\caption{Real (solid lines) and imaginary (dashed lines) parts of the parallel (red lines) and perpendicular (blue lines) dielectric functions Si:InAs/AlSb HMM ($c_{\bf 2}=$3.12\%)  with equal filling factors ($f_1=0.5$), as a function of wavelength (instead of energy as in Figure 3 of the main text).}
\label{compexp2}
\end{figure*}

\begin{figure*}[]
\centering
\includegraphics[width=0.85\textwidth]{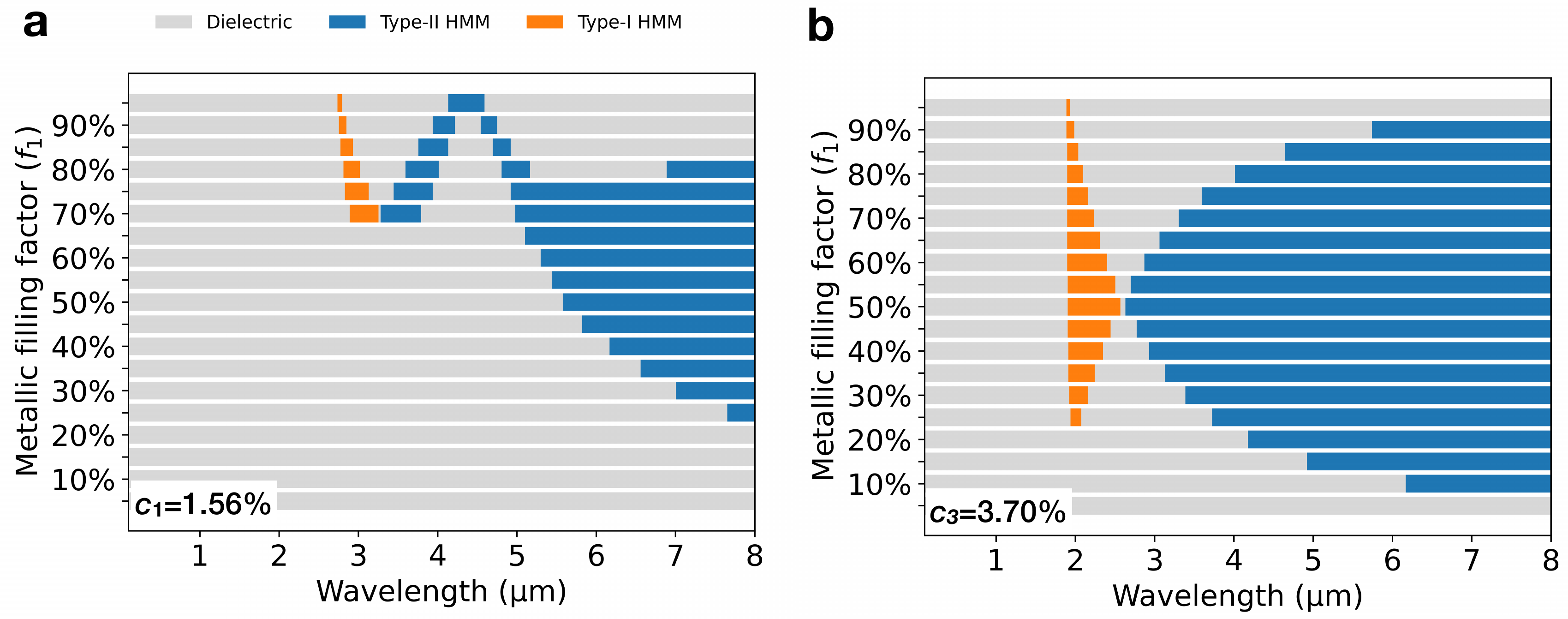}
\caption{Spectral optical character -- namely Type-I (orange) and Type-II (blue) hyperbolic, and dielectric (gray) --  of 
    Si:InAs/AlSb HMMs  for (a) the lowest $c_{\bf 1}=1.56\%$  and (b) highest  $c_{\bf 3}=3.70\%$  doping concentrations, as a function of the metallic filling factor $f_m$ in the range (0.05 -- 0.95)\%. Intermediate doping case $c_{\bf 2}=3.12\%$ is reported in Figure 4c of the main text.}
\label{doping_filling}
\end{figure*}

\begin{figure*}[]
\centering
\includegraphics[width=0.6\linewidth]{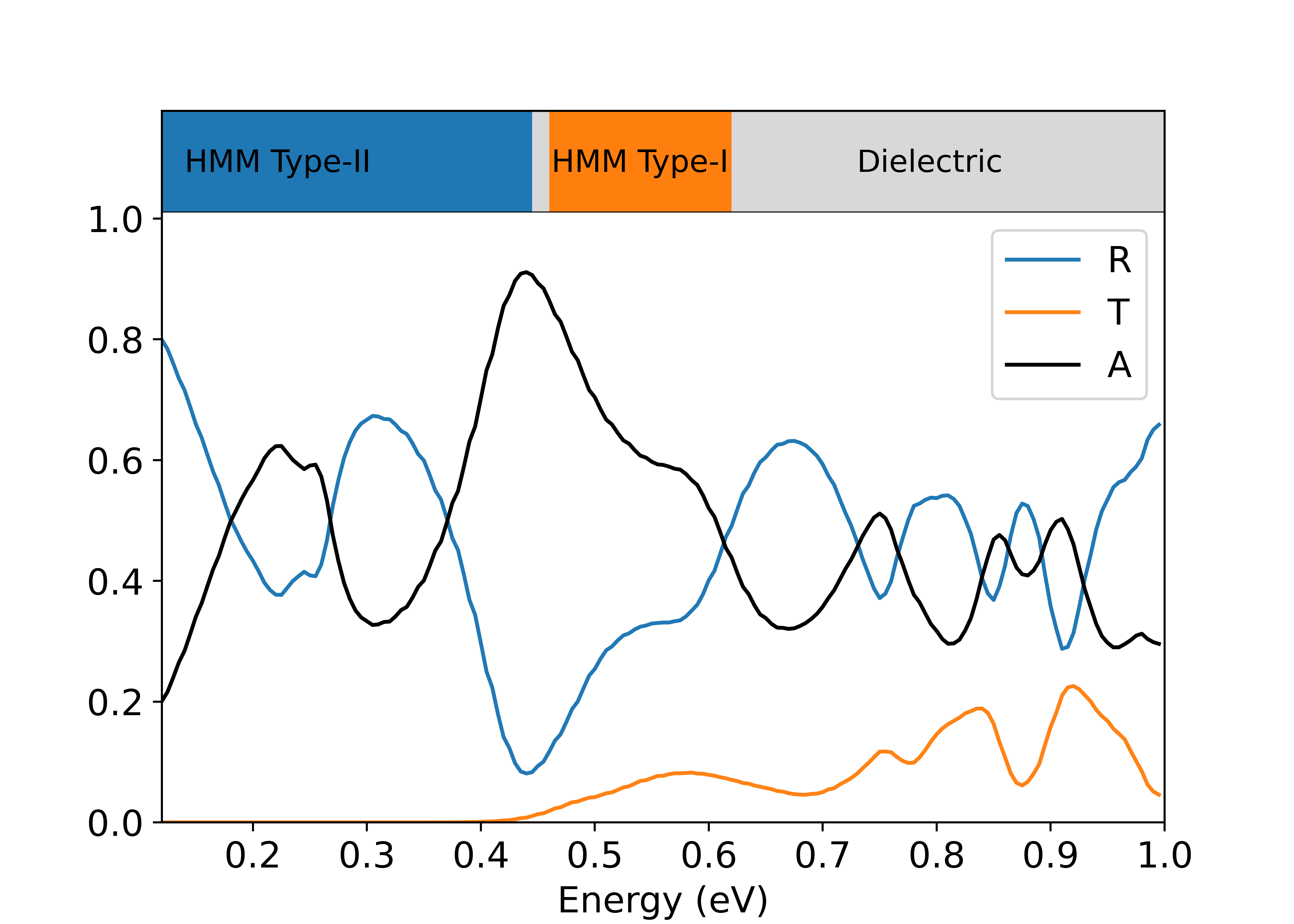}
\caption{Reflectivity (blue), transmittance (orange) and absorption (black) spectra for Si:InAs/AlSb multilayers ($c_{\bf 2}=3.12\%$, $f_m=0.5$, $\mathcal{N}=10$). Symbols and labels refer to main text.}
\label{trasmission}
\end{figure*}

    \end{document}